\documentclass[12pt]{article}
\usepackage[margin=1in]{geometry}
\usepackage[utf8]{inputenc}
\usepackage[T1]{fontenc}

\usepackage[table]{xcolor}
\definecolor{lightblue}{RGB}{220,235,255}

\usepackage{tikz}
\usetikzlibrary{arrows.meta, positioning, calc, fit, shapes.geometric}

\usepackage{amsmath,amssymb,amsfonts}
\usepackage{algpseudocode}
\usepackage{algorithm}
\usepackage{graphicx}
\usepackage{textcomp}
\usepackage{booktabs}
\usepackage[english]{babel}
\usepackage{csquotes}
\usepackage[style=apa, backend=biber]{biblatex}
\usepackage{multirow}
\usepackage{placeins}
\usepackage{siunitx}
\usepackage{caption}
\usepackage[normalem]{ulem}

\usepackage{hyperref}
\usepackage{url}

\usepackage{endnotes}
\let\footnote=\endnote

\usepackage{xcolor}
\usepackage{soul}
\sethlcolor{yellow}

\makeatletter
\renewcommand\footnotesize{%
   \@setfontsize\footnotesize{7pt}{10pt}%
}
\makeatother

\begin{document}

\title{Improving Access to Historical Archives with Real-time RAG-based Systems}

\author{%
\begin{tabular}[t]{@{}c@{\hspace{2em}}c@{\hspace{2em}}c@{}}
\begin{tabular}[t]{@{}c@{}}
Stergios Konstantinidis \\
University of Lausanne \\
Lausanne, Switzerland \\
stergios@unil.ch \\
\textit{Corresponding author}\\
\end{tabular}
&
\begin{tabular}[t]{@{}c@{}}
Hayman Lotfy \\
University of Lausanne \\
Lausanne, Switzerland \\
hayman.lotfy@unil.ch
\end{tabular}
&
\begin{tabular}[t]{@{}c@{}}
Alexis Erne \\
University of Lausanne \\
Lausanne, Switzerland \\
alexis.erne@ik.me
\end{tabular}
\\[2em]
\begin{tabular}[t]{@{}c@{}}
Faruk Zahiragic \\
Swiss Federal Institute \\
of Technology (EPFL) \\
Lausanne, Switzerland \\
faruk.zahiragic@epfl.ch
\end{tabular}
&
\begin{tabular}[t]{@{}c@{}}
Min-Yen Kan \\
National University \\
of Singapore (NUS) \\
knmnyn@nus.edu.sg
\end{tabular}
&
\begin{tabular}[t]{@{}c@{}}
Michalis Vlachos \\
University of Lausanne \\
Lausanne, Switzerland \\
michalis.vlachos@unil.ch
\end{tabular}
\end{tabular}%
}
\date{}
\maketitle

%TC:ignore
\begin{abstract}
Digitized historical archives are large, heterogeneous cultural heritage repositories, but access methods for such archives face challenges such as noisy optical character recognition (OCR) output and rigid keyword-based retrieval, which limit retrieval quality. In this work, we present an end-to-end archival processing and retrieval framework that integrates large language models (LLMs) into the archival pipeline. Our system introduces two core components: (i) an LLM-based OCR refinement module that improves text quality, and (ii) a semantic retrieval and cross-encoder reranking pipeline supporting natural-language question answering via retrieval-augmented generation (RAG). Our evaluations are done on a historical archival dataset of 500,000 Swiss newspaper segments spanning over three centuries (1762–2001). Experiments are conducted across 384 natural-language test queries. Our results highlight that LLM refinements reduce OCR errors by up to 44.52\% (CER) and 60.95\% (WER). More importantly, this is accompanied by downstream information retrieval improvements. Compared to traditional keyword baselines, our reranking pipeline increases NDCG@10 by 31.9\% (from 65.99\% to 87.05\%) and achieves statistically significant gains in both answer correctness and context relevance. These results demonstrate that integrating LLMs with established document processing and retrieval pipelines can elevate digital libraries from static repositories to interactive, semantically searchable archival systems.
\end{abstract}
%TC:endignore

%TC:ignore

\section{Introduction}

Digitized historical archives represent some of the most important publicly accessible cultural heritage collections. National, regional, and university libraries have invested heavily in the digitization of newspapers, magazines, and other printed material spanning several centuries. In most cases, these collections are made available through a combination of scanned page images and OCR-derived text, together with search interfaces intended to facilitate access to their contents. The Cantonal and University Library of Lausanne (\textit{Bibliothèque Cantonale et Universitaire de Lausanne}, BCUL), for example, maintains a major digitized newspaper archive with issues dating back to the eighteenth century. Such repositories are invaluable for historians, journalists, students, and the wider public. However, despite their scale and documentary value, these repositories remain underused because their contents are not always easily accessible in practice \autocite{kumpulainen2022struggling}.

A major reason for this limitation is that archival access still depends on fragile textual representations and rigid retrieval paradigms. First, OCR applied to historical documents often introduces substantial transcription errors. These errors arise from multiple sources, including document degradation, low-quality scans, typographic variability, unusual page layouts, and the graphical complexity of older print traditions. As a result, the OCR text stored alongside digitized pages is frequently noisy and incomplete. This affects not only readability, but also every downstream component that relies on textual input, including indexing, retrieval, and automated analysis.

Second, the dominant access model for archival platforms is still based on keyword search. While this approach is adequate for many contemporary corpora, it is significantly less effective for historical collections. Exact lexical matching is brittle when spelling conventions shift across time, when users do not know the specific historical terminology used in the source material, or when relevant content is expressed through paraphrase rather than direct lexical overlap. In practice, many relevant documents remain undiscovered simply because the query language and the archival language do not align. As a result, valuable archival content may remain effectively hidden even when it has already been digitized and indexed.

Recent progress in large language models (LLMs) and retrieval-augmented generation (RAG) provides an opportunity to revisit this problem from a new perspective. RAG systems combine retrieval with language generation, allowing users to formulate information needs in natural language and receive synthesized answers grounded in supporting documents. This interaction model is particularly appealing for archival collections, since it reduces the dependence on exact keyword matching and opens the possibility of more intuitive exploration. However, applying RAG to noisy historical corpora is not straightforward. OCR errors degrade the quality of the indexed text, which in turn can harm dense retrieval, reranking, and answer generation. Moreover, many modern RAG pipelines are computationally heavy. Their multi-stage architecture—typically involving embedding-based retrieval, cross-encoder reranking, and autoregressive response generation—can lead to non-negligible latency, making them poorly suited for interactive exploratory use.

These challenges suggest that archival access should not be framed solely as a search problem, but more broadly as a data systems problem involving ingestion, representation, indexing, retrieval, and interactive serving. From an information science perspective, this study examines how noisy cultural heritage records can be transformed into more reliable information representations, and how these representations affect retrieval, evidence inspection, and human access to archival knowledge. In this work, we propose an end-to-end architecture for historical newspaper exploration that explicitly accounts for this full pipeline. Our design treats the archive as a managed data resource whose usefulness depends both on the quality of the textual representation produced at ingestion time and on the efficiency and robustness of the retrieval workflow applied at query time.

At ingestion time, we use instruction-tuned LLMs as post-processing operators for OCR output. Rather than treating OCR as a fixed preprocessing step, we regard it as an intermediate representation that can be refined before indexing. The goal is not to rewrite archival content, but to improve transcription fidelity while preserving structure and semantic faithfulness. By improving the textual representation used to build the search index, this refinement stage benefits the entire downstream system.

At query time, we implement a semantic retrieval architecture that combines dense retrieval, reranking, and grounded answer generation. To support practical interaction over large archival collections, the system is designed with latency constraints in mind. We rely on approximate nearest-neighbor indexing to make retrieval scalable, and we study the trade-off between reranking quality and computational overhead. In addition, we adopt streaming generation so that answers can be returned incrementally, reducing perceived latency and making the system more appropriate for interactive exploration scenarios. In our deployed setting, this allows response times on the order of a second while preserving meaningful retrieval quality.

We evaluate the proposed pipeline using real historical newspaper data and study its behavior along three complementary dimensions. First, we assess whether LLM-based post-OCR refinement improves the quality of noisy archival text. Second, we examine whether these improvements propagate to downstream retrieval and answer generation. Third, we analyze system-level trade-offs between retrieval effectiveness and response time in order to identify configurations suitable for interactive use. Beyond quantitative evaluation, we also demonstrate the practical applicability of the approach through a user-facing interface for immersive archive exploration.

\section{Research Objectives and Contributions}
Our core objectives can be summarized as such:

\begin{itemize}
    \item \textbf{Overcoming Ingestion Noise via LLM Refinement.} Our first objective is to evaluate whether instruction-tuned Large Language Models (LLMs) can effectively sanitize noisy OCR text to create a higher-quality foundational corpus. To this end, we develop an ingestion pipeline that uses LLMs as post-OCR refinement operators for historical archival text. Our results demonstrate that the effectiveness of this sanitization depends heavily on model scale and instruction-following ability, with larger models producing more reliable transcription improvements.
    
    \item \textbf{Enabling Semantic Retrieval through RAG.} To measure the effectiveness of replacing traditional exact-match keyword systems, we evaluate a Retrieval-Augmented Generation (RAG) architecture's impact on answer correctness and context relevancy. As a contribution, we design a semantic retrieval architecture combining dense retrieval, cross-encoder reranking, and grounded answer generation. We show that improving OCR quality at ingestion time leads to measurable, statistically significant gains in downstream retrieval effectiveness.
    
    \item \textbf{Facilitating Interactive Access with Latency-Aware Processing.} Ensuring that real-time exploration of large-scale collections is practically deployable requires balancing retrieval accuracy with computational efficiency. We contribute a latency-aware RAG pipeline that achieves this balance through approximate nearest-neighbor search, lightweight reranking, and streaming generation, successfully enabling interactive question answering over large archives.
    
    \item \textbf{Demonstrating Practical Relevance through a Prototype Interface.} Building upon the previous objectives, we implement a prototype user-facing interface that supports natural-language exploration of archival material. This application allows users to inspect retrieved supporting evidence, demonstrating the real-world viability of our approach.
    \end{itemize}

\section{Related Work}

In recent years, the integration of information systems into digital archival libraries has transformed how historical documents are indexed, retrieved, and consumed. For example, the Library of Congress in the United States holds more than 16 million digital newspaper pages through its ``Chronicling America'' initiatives, spanning from the 17th to the 20th century \autocite{lee2020newspaper}. Other notable efforts include Scriptorium\footnote{\url{https://www.scriptorium.ch}}, the digital portal of the Cantonal and University Library of Lausanne with more than 7.5 million scanned pages. Both efforts exemplify the traditional pipeline of OCR-based transcription followed by basic keyword search, which remains a common starting point for research on digitized historical newspapers \autocite{oberbichler2022integrated}. However, such traditional access methods face major limitations due to OCR noise and restricted search capabilities. Emerging research addresses these challenges by leveraging large language models and retrieval-augmented generation to support more semantic and accessible interactions with archival data \autocite{song2026transforming}. Similarly, in other domains dealing with long and structurally complex documents, recent frameworks like H-ProtoRAG demonstrate that combining hybrid retrieval with cross-encoder reranking can successfully overcome the limitations of traditional keyword-based search \autocite{ELMI2026104861}. In the following sections, we review how research on LLMs and related technologies is helping to advance archival search capabilities.

\medskip
\noindent
\textbf{LLMs for Data Sanitization.}
Most data derived from digitized historical documents contain transcription errors, often stemming from complex typographic features such as elongated characters \autocite{vesalainen_error_2026}, which are then propagated to downstream tasks such as RAG \autocite{zhang2025ocr}. Although manual corrections ensure high precision, automated methods are essential to sanitize large datasets. LLMs, with their broad multilingual knowledge, represent promising tools for addressing OCR-related challenges \autocite{chowdhery2023}. Only very recent and preliminary work uses LLMs for post-OCR correction \autocite{barman2021,thomas2024,manrique2024historical,boros_ehrmann2024,do2025,levchenko2025evaluating,10.1117/12.3104502}. However, the results are conflicting, with some reporting improvements and others deterioration. These discrepancies can largely be explained by differences in datasets (e.g., historical vs. modern text; varying OCR noise profiles), evaluation criteria (lexical fidelity vs. semantic coherence) modeling choices (zero-shot prompting vs. fine-tuning, context length) and LLM over-correction. To mitigate those shortcomings and reduce LLM usage and cost, strategies such as selective correctness have been elaborated \autocite{10.1117/12.3068186}. Our results on a novel and diverse dataset show that fine-tuned or instruction-tuned LLMs \autocite{zhao2023survey} can improve, but not completely eliminate, OCR-induced errors. Beyond archival text, the robust capacity of deep learning models to handle highly complex and noisy linguistic structures has been increasingly validated in other fields. For example, recent advances use hierarchical and multi-scale architectures to successfully process noisy clinical records \autocite{RUAN2026104784} and morphologically rich texts for automated grading  \autocite{Zhu_et_al}. These cross-domain successes further justify our application of LLMs as sanitization operators. More importantly, to the best of our knowledge, this is among the first study demonstrating that post-OCR corrections performed by LLMs also improve performance within a RAG framework.

\medskip
\noindent
\textbf{Interactive and Immersive Interfaces.}
While the backend processing of archival data has advanced, the front-end user experience has largely remained confined to 2D web-based viewers. Recent efforts have begun exploring more immersive modalities to increase user engagement with search interfaces \autocite{10.1145/3539618.3592057}. Other efforts have centered on studying the Computer-Human interactions in an augmented reality environment, specifically with historical and archival documents \autocite{10.1145/3706599.3720092}.

\medskip
\noindent
\textbf{AI Platforms for Digital Libraries.}
Recent years have witnessed the emergence of AI-driven platforms designed to enhance archival discovery, navigation, and interpretation. Several institutions have begun integrating large language models into digital library workflows, replacing traditional OCR and keyword-based systems with semantically rich, context-aware search and summarization capabilities.

One prominent example is deployed at Virginia Tech's Digital Library \autocite{banerjee2024making}, which applies LLMs to process scanned historical materials, including handwritten letters and maps. Their system combines handwriting recognition, natural language summarization, and entity extraction to generate searchable text representations and user-friendly summaries, significantly improving the accessibility of historical collections. However, their solution does not provide a natural language search paradigm as we do in this work, nor do they use any metric indexing technique for fast search as we propose.

Transkribus, another widely adopted platform for historical document processing, leverages handwritten text recognition (HTR) techniques for transcription \autocite{transkribus2022}, which can then be post-processed by LLMs such as ChatGPT for normalization and interpretation. Recent work \autocite{spina2023archival} has evaluated this combined HTR+LLM pipeline on the Biscari Archive (366 handwritten letters from World War II), showing that while LLMs improve post-OCR legibility and structure, they still require human oversight to preserve historical accuracy.

Other AI platforms have focused on metadata creation and enrichment. Boros et al. \autocite{boros_ehrmann2024} demonstrate how LLMs can automate archival description tasks, generate multilingual metadata, and surface underrepresented content by analyzing document structure and semantics. Jaillant and Rees \autocite{jaillant2022applying} emphasize the need for trust and collaboration between archivists and technologists to ensure ethical and inclusive AI applications in heritage contexts. Similarly, Manrique-Gómez et al.~\autocite{manrique2024historical} developed a corpus of 19th-century Latin American newspapers with OCR corrections driven by LLMs, highlighting their effectiveness in handling noisy historical Spanish-language content.

More advanced systems, such as the federated LLM system described by Groppe et al.~\autocite{groppe2025automated}, coordinate multiple large language models with validation and synthesis to automatically generate archival metadata that conforms to standards across heterogeneous document types and formats. This work exemplifies a broader move from static keyword workflows toward AI-assisted archival processing focused on consistent, high-quality metadata, rather than on search interfaces. Such complex pipelines are justified in high-stakes projects, but may be too expensive to manage for resource-constrained digitization projects. Furthermore, as digital libraries increasingly rely on autonomous LLM agents for these critical curation tasks, integrating structured verification and memory mechanisms—such as those proposed in the Veritas framework—becomes essential to ensure logical consistency and mitigate LLM hallucinations \autocite{TANG2026104830}.

Our work expands upon and integrates these threads by developing an end-to-end platform that jointly optimizes OCR correction and RAG-based semantic search. In contrast to index-driven approaches such as \autocite{ehrmanncomputational}, which leverage historical indexes to bootstrap access and guide digitization and align multiple indexes into a single meta-index for referential retrieval, our pipeline spans the full lifecycle, from raw scan to conversational query, enabling nuanced access to complex historical collections. Using a curated dataset of actual newspaper archives spanning more than two centuries, we demonstrate measurable reductions in OCR error rates and significant improvements in contextual search quality. 

\begin{figure}[!t]
    \centering
    \includegraphics[width=0.99\linewidth]{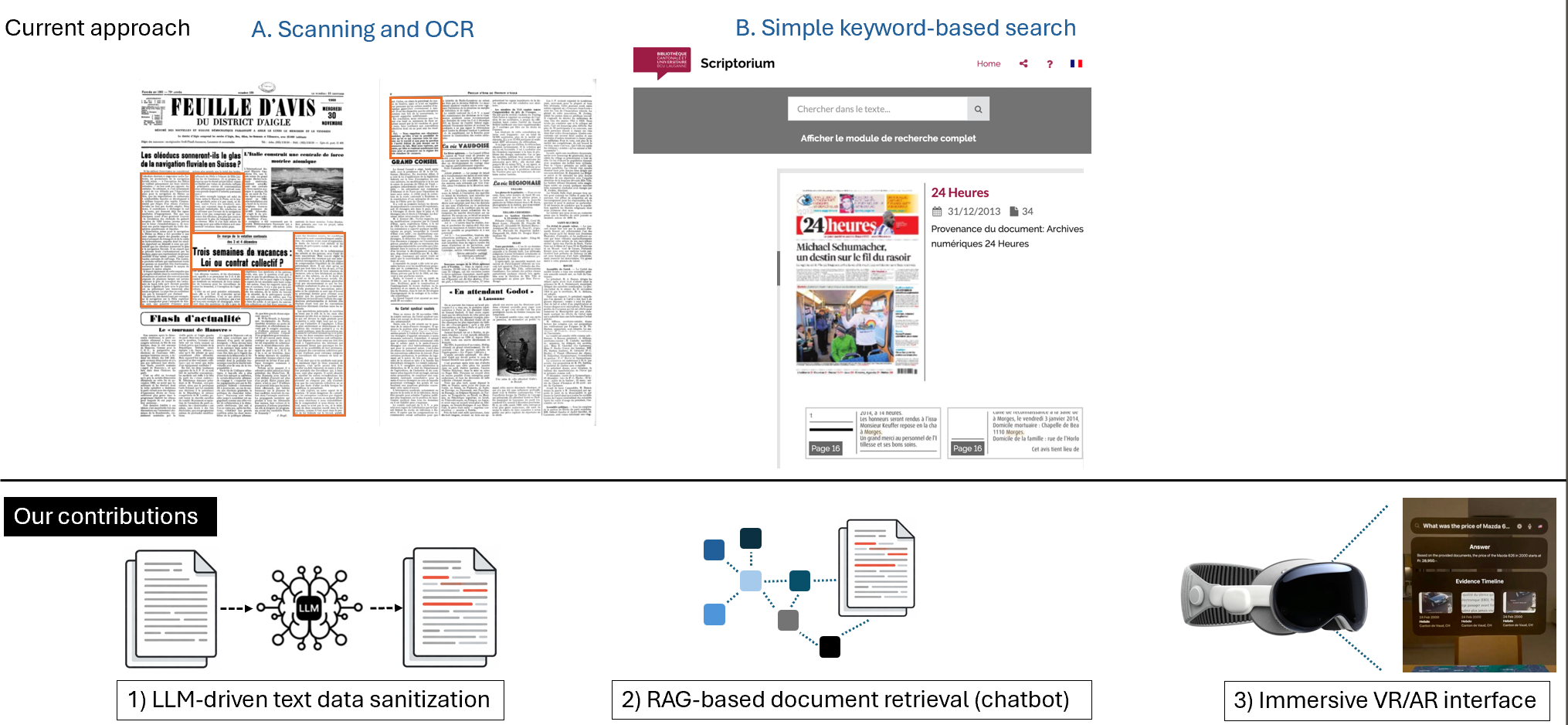}
    \caption{Top: Typical archival access pipelines rely primarily on (A) OCR and (B) keyword-based search. Bottom: Our contributions: 1) we use LLMs to correct OCR errors, 2) support natural language queries over the historical archive, and 3) create new AR/VR interfaces.} \label{fig:overview}
\end{figure}

This unified approach represents one of the first deployments of instruction-tuned LLMs at scale for a real-world digital library dataset, addressing both technical challenges and curatorial needs in historical document accessibility. Figure~\ref{fig:overview} illustrates the difference between current implementations and our approach.

\section{Methodology}

First, we describe an LLM-assisted pipeline for OCR error correction. We then augment it with a retrieval-augmented generation framework that supports natural language querying over the corrected document collection. We evaluate whether RAG-based retrieval \autocite{fan2024survey} enables semantic, context-aware access to historical archives, moving beyond traditional keyword-based search by retrieving relevant documents based on meaning rather than exact lexical overlap. Quantitative evaluation demonstrates measurable improvements in information retrieval metrics, including NDCG, compared to classical IR baselines. By indexing archival documents using vector representations, the proposed system facilitates intent-aware retrieval that better captures semantic relevance and historical context.

To support efficient retrieval over large archival corpora, we index document embeddings using a graph-based approximate nearest-neighbor structure. In particular, we employ a hierarchical graph index that enables fast exploration of the vector space while preserving high retrieval quality. This structure allows sublinear query time in practice and produces an initial candidate set of documents for further refinement.

This initial retrieval stage is optimized for recall and speed, producing a broad candidate set of potentially relevant documents. To improve precision, we subsequently apply a cross-encoder reranking model that jointly evaluates each query--document pair. This two-stage design decouples fast approximate retrieval from more computationally intensive relevance scoring, enabling scalable yet accurate semantic search over large collections.

\begin{figure*}[!ht]
    \centering
    \includegraphics[width=0.99\linewidth]{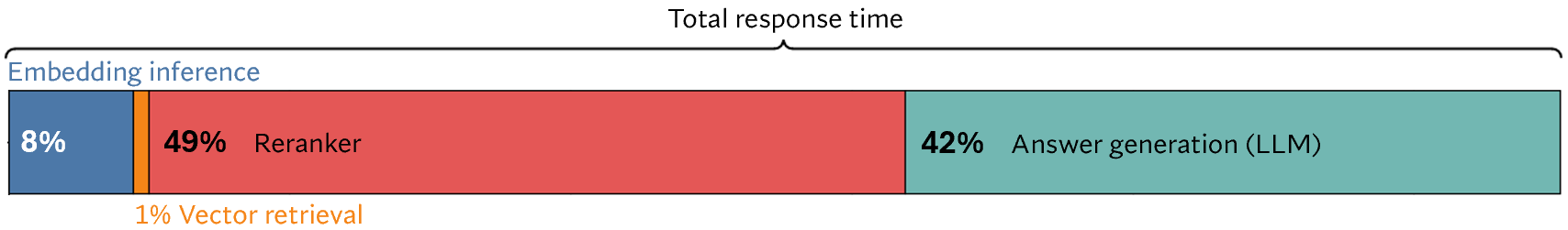}

    \caption{Indicative allocation of computation time across the stages of a RAG pipeline.}
    \label{fig:timing_breakdown}
\end{figure*}

\medskip
\noindent
\textbf{Latency-aware design.}
A key requirement for interactive archival systems is low response time. In our pipeline, the main computational cost arises from reranking and answer generation, as shown in Figure~\ref{fig:timing_breakdown}, while vector retrieval remains comparatively inexpensive. Increasing the number of candidate documents improves retrieval quality but also increases reranking cost, leading to a trade-off between effectiveness and latency.

To address this, we adopt a lightweight reranking strategy and combine it with streaming generation during inference. Instead of waiting for full completion, the system progressively returns tokens as they are generated, allowing users to start consuming the response immediately. This design improves perceived responsiveness and makes the system suitable for interactive exploration scenarios.

\medskip
\noindent
\textbf{Sanitizing OCR Errors.}
Let $\mathcal{D} = \{D_1, D_2, \ldots, D_n\}$ denote the collection of digitized issues provided in PDF format by a digital library provider. Each issue $D_i$ is segmented into a set of articles
\[
A(D_i) = \{a_{i1}, a_{i2}, \ldots, a_{im_i}\}
\]
using the \textsc{LayoutParser} framework \autocite{shen_layoutparser:_2021,doclaynet2022}, which enables structured extraction of text regions from scanned documents.

For each article $a_{ij}$, we apply OCR to obtain a machine-readable text string:
\[
T_{ij} = \mathrm{OCR}(a_{ij}),
\]
where $T_{ij}$ may contain transcription errors due to noise in the original scans.

To mitigate OCR-related errors and improve text quality, we employ large language models as post-processing functions. Specifically, let
\[
\hat{T}_{ij} = \mathrm{LLM}_{\theta}(T_{ij}, P, C)
\]
denote the corrected text, where $P$ represents an engineered prompt and $C$ denotes an augmented context provided to the model. Following \autocite{sahoo_systematic_2025}, we leverage both strategies to modify $P$ and $C$: (i) enhancing the prompt $P$ accompanying the input text, and (ii) adjusting the context $C$ with task-specific guidance.

The design ensures that $\hat{T}_{ij}$ both preserves the structure of $T_{ij}$ and conforms to the formatting constraints of the original article.

The standardized contexts $C$ include worked examples, task-specific formatting instructions, and explicit length constraints, which collectively ensure consistency of model outputs even under reduced context windows. The result is a dataset that is both semantically faithful and structurally aligned with the original digitized sources.

\medskip
\noindent
\textbf{RAG on the Sanitized OCR Texts.}
We outline a retrieval-augmented generation framework that enables flexible natural language search across both sanitized and non-sanitized versions of the corpus. Let $q$ denote a user query, and let $\phi$ be an embedding function. For each article $a_{ij} \in A(D_i)$ we define two textual variants: the raw OCR text $T_{ij}$ and its LLM-corrected counterpart $\hat{T}_{ij}$. We obtain corresponding embeddings
\[
\mathbf{z}_{ij} = \phi\!\big(T_{ij}\big), 
\qquad 
\hat{\mathbf{z}}_{ij} = \phi\!\big(\hat{T}_{ij}\big),
\]
and a query embedding $\mathbf{r}=\phi(q)$. We measure semantic proximity using the cosine distance
\[
d_{\cos}(\mathbf{u},\mathbf{v}) = 1 - \frac{\mathbf{u}^{\top}\mathbf{v}}{\|\mathbf{u}\|_2\,\|\mathbf{v}\|_2}
\quad\in[0,2].
\]

For a given corpus view $X\in\{\text{raw},\text{sanitized}\}$ with embeddings
\[
\mathbf{z}^{(X)}_{ij} =
\begin{cases}
\mathbf{z}_{ij}, & X=\text{raw},\\
\hat{\mathbf{z}}_{ij}, & X=\text{sanitized},
\end{cases}
\]
we retrieve the top-$k$ nearest articles to $q$ by minimizing cosine distance (equivalently, maximizing cosine similarity):
\[
\mathcal{N}_k(q;X) = \operatorname{TopK}_k\!\Big\{ -\,d_{\cos}\!\big(\mathbf{r},\mathbf{z}^{(X)}_{ij}\big) : a_{ij}\in A(D_i),\, \forall i,j \Big\}.
\]

The retrieved set induces a context
\[
\mathcal{C}(q;X) = \big\{ X_{ij} : a_{ij}\in \mathcal{N}_k(q;X)\big\},
\]
where $X_{ij} \in \{T_{ij},\hat{T}_{ij}\}$ matches $X$.

An LLM then generates a grounded answer conditioned on the query and its retrieved context:
\[
\hat{Y}(q;X) = \mathrm{LLM}_{\theta}\!\big(q,\; \mathcal{C}(q;X)\big).
\]

This formulation allows systematic and comparable evaluation across embedding functions $\phi$ and corpus views $X$, isolating the effect of OCR sanitization on both retrieval quality and answer quality.

To enable scalable retrieval over large archival collections, embeddings $\mathbf{z}^{(X)}_{ij}$ are indexed using an approximate nearest neighbor search structure \autocite{johnson2019billion}, providing sublinear query time while maintaining high recall. This ANN-based retrieval supplies the initial candidate set for reranking and downstream generation.

\medskip
To improve retrieval precision, we further apply a reranking stage over the initial candidate set $\mathcal{N}_k(q;X)$. Let $\psi$ denote a cross-encoder reranker that assigns relevance scores $s_{ij}=\psi(q,X_{ij})$ to each retrieved article. The final context is constructed from the top-$k_r$ reranked items,
\[
\mathcal{N}_{k_r}^{\text{rerank}}(q;X) = \operatorname{TopK}_{k_r}\{ s_{ij} : a_{ij}\in \mathcal{N}_k(q;X)\},
\]
which replaces $\mathcal{N}_k(q;X)$ in the construction of $\mathcal{C}(q;X)$ during evaluation.

Finally, given the top-$k_r$ retrieved and reranked articles, the resulting context $\mathcal{C}(q;X)$ is provided to a pretrained large language model, which generates a natural language response conditioned jointly on the user query $q$ and the retrieved evidence. This generation step follows standard retrieval-augmented generation practice, producing answers grounded in the selected archival content.

\medskip \noindent \textbf{Response time:}
Response time is a critical requirement for interactive systems. Figure~\ref{fig:timing_breakdown} reports the mean total response time across the stages of a RAG pipeline (additional details are provided in the experimental section). The breakdown indicates that reranking and answer generation dominate the overall latency, suggesting that these stages are the primary targets for optimization.

The reranking stage jointly evaluates each query--candidate pair to compute a refined relevance score, but introduces an $O(L \cdot d^2)$ computational bottleneck, where $L$ denotes the reranker candidate pool size and $d$ the sequence length (in tokens) of the query--candidate pair. In practice, this creates a trade-off between retrieval quality and latency: increasing $L$ may improve information retrieval metrics (e.g., recall or NDCG), but also increases computational cost. Practitioners therefore need to select a suitable candidate pool size that balances effectiveness and response time.

Answer generation also contributes significantly to end-to-end latency. In our implementation, we use streaming generation from the LLM that returns tokens incrementally. This approach improves perceived responsiveness, as partial outputs become visible while generation is still in progress. Since users can begin reading the response as it is produced, streaming overlaps generation time with user interaction. We further examine this behavior in the experimental evaluation.

\section{Experiments}

Our empirical evaluation aims to address the following research questions:

\begin{enumerate}
\item \uline{Do LLM sanitizations improve the quality of raw OCR text?} Prior studies have reported mixed findings, with some observing improvements and others reporting degradation. Our experiments allow us to systematically assess this effect and reveal that smaller language models tend to degrade text quality, whereas larger and more recent models lead to measurable improvements.

\item \uline{Do improvements in OCR text quality translate into downstream information retrieval benefits?} Using a real-world archival library dataset, we evaluate whether gains at the text level propagate to retrieval performance. To the best of our knowledge, this is the first work on historical archives that systematically measures this effect, as all previous studies \autocite{boros_ehrmann2024,do2025,levchenko2025evaluating,manrique2024historical} focused solely on LLM refinements over the raw OCR text.

\item \uline{How does the proposed system balance retrieval quality and response time?} We analyze the trade-off between candidate pool size, reranking complexity, and generation latency to identify configurations suitable for interactive use.
\end{enumerate}

Beyond quantitative evaluation, we also present qualitative visual examples that illustrate the impact of OCR refinement on retrieval and RAG-based search.

\begin{figure*}[!ht]
    \centering
    \includegraphics[width=0.9\linewidth]{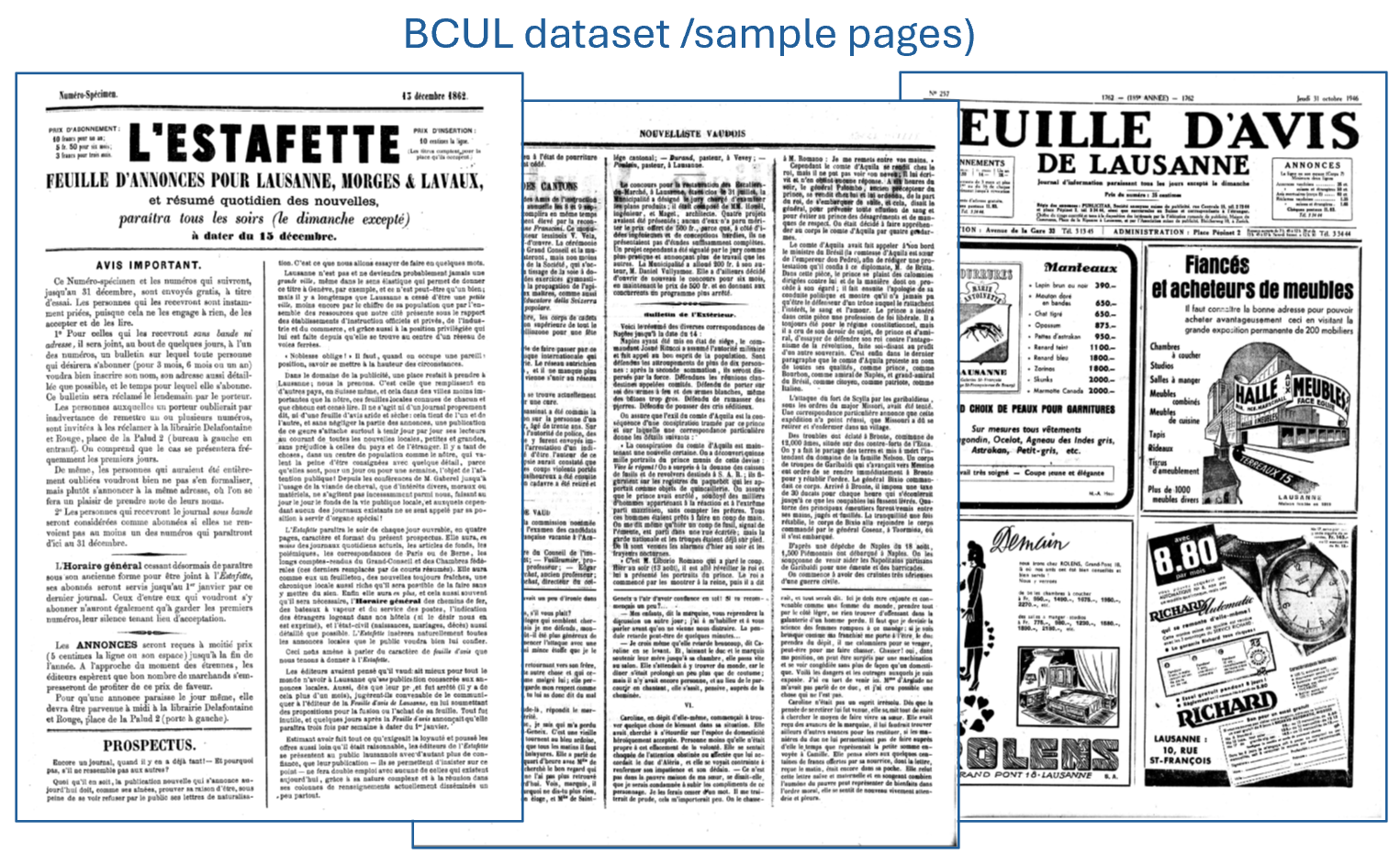}
    \caption{Examples of the datasets used.}
    \label{fig:dataset_samples}
\end{figure*}

\medskip
\noindent
\textbf{Datasets.}
Our experimental evaluation was conducted on a curated subset of historical newspaper issues provided by the Cantonal Library of Lausanne via the Scriptorium portal. This is a novel and highly diverse dataset. Because this dataset has not been previously published, its curation and semantic enrichment represent an independent contribution of this work. This dataset was compiled from nine distinct periodicals, spanning 239 years, from 1762 to 2001. The temporal coverage includes near-continuous annual representation from the late 18th century through 1913, a substantial block of mid-20th-century issues (1950–1959), and selective modern samples up to 2001. 
From this vast corpus, we constructed a representative evaluation index of 500,000 randomly sampled document segments. Crucially, to support complex Retrieval-Augmented Generation (RAG) evaluations, these segments are not merely raw OCR text chunks. Each segment in our index is enriched with highly structured metadata, including:

Source \& Spatial Tracking: Issue IDs, page numbers, and exact bounding-box coordinates anchoring the text to the original scanned image.
Temporal Markers: Exact publication date to enable chronological filtering.
Semantic Enhancements: AI-generated abstractions for each segment, including a synthesized title, a concise summary, and extracted keywords.
Named Entity Recognition (NER): Categorized lists of entities mentioned in the text, such as Persons, Locations, Organizations, and Events.
This rich metadata structure allows our system to perform hybrid search strategies—combining dense vector retrieval with hard metadata filtering. Alongside this corpus, we created an evaluation set of 384 natural language queries. These queries were manually formulated and include five adversarial "unanswerable" questions designed specifically to test the system's susceptibility to hallucination.
Figure~\ref{fig:dataset_samples} illustrates some samples from our dataset.

\medskip
\noindent
\textbf{OCR Engines.}
We benchmarked 6 OCR systems spanning open-source toolkits, paid APIs, and on-device frameworks:

\begin{itemize}
  \item \textit{Tesseract} is one of the most widely used open-source OCR engines \autocite{33418}.
  \item \textit{Docling} is a modern document-conversion pipeline that integrates layout analysis with OCR to produce structured text suitable for downstream processing \autocite{docling}.
  \item \textit{EasyOCR} leverages deep neural networks for text detection and recognition across a wide range of alphabets and scripts \autocite{easyocr}.
  \item \textit{Apple Vision OCR} offers on-device recognition tightly integrated with the Vision framework.\footnote{\url{https://developer.apple.com/documentation/vision/vnrecognizetextrequest}} It provides a modern multi-stage neural pipeline, combining deep text detection with sequence-based recognition and language-aware decoding. This end-to-end design typically yields higher robustness to noise, blur, and real-world capture conditions, as seen in our experiments.
  \item \textit{Azure Document Intelligence} is a cloud-based document processing service that combines optical character recognition with layout analysis and structured information extraction. It leverages deep learning models trained on large-scale document corpora to produce high-quality text representations and metadata, and is designed to handle complex document layouts and noisy inputs commonly found in real-world scenarios.

  \item \textit{Google Document AI} provides an end-to-end document understanding platform that integrates OCR with advanced layout parsing and semantic structure extraction. Its models are optimized for diverse document types and support robust text recognition under challenging conditions, including degraded scans and heterogeneous formatting, making it well-suited for large-scale archival processing.

\end{itemize}

\medskip
\noindent
\textbf{Large Language Models.}
We evaluated a range of instruction-tuned LLMs for two tasks: (i) post-processing OCR text to reduce transcription errors, and (ii) powering retrieval-augmented generation for semantic search. The models span open-source baselines, domain-specific models, and frontier commercial systems:

\begin{itemize}
  \item \textit{Gemma 3} \autocite{gemma_2025} serves as a compact open-source model, effective for lightweight OCR correction but less consistent for complex historical inputs.
  \item \textit{Qwen3 (235B)} \autocite{qwen3technicalreport} represents a large-scale open-source model trained on multilingual corpora, showing strong capacity for OCR correction but variable stability across benchmarks.
  \item \textit{DeepSeek-r1} \autocite{deepseekai2025deepseekr1incentivizingreasoningcapability} is one of the biggest open-source models, however it only demonstrated moderate improvements.
  \item \textit{GPT-4o, GPT-5 and GPT-5.2} \autocite{openai_gpt-4_2024,openai_gpt5_2025,openai_gpt52_2025} offered robust instruction-following and stable editing behavior, delivering consistent improvements in both OCR correction (WER and CER) and RAG-based correctness metrics.
  \item \textit{Gemini 2.5 Pro, 3 Pro and 3 Flash} \autocite{comanici_gemini_2025,gemini3pro,gemini3flash}, the most advanced closed-source models in our study, achieved the strongest overall gains, substantially reducing error rates in OCR correction and yielding the highest relevancy scores in RAG evaluations.
\end{itemize}

Our results on these language models suggest that while open-source models provide a useful baseline, large commercial LLMs deliver the most reliable improvements for both OCR sanitization and semantically grounded retrieval.

\begin{figure}[!ht]
    \centering
    \includegraphics[width=.8\linewidth]{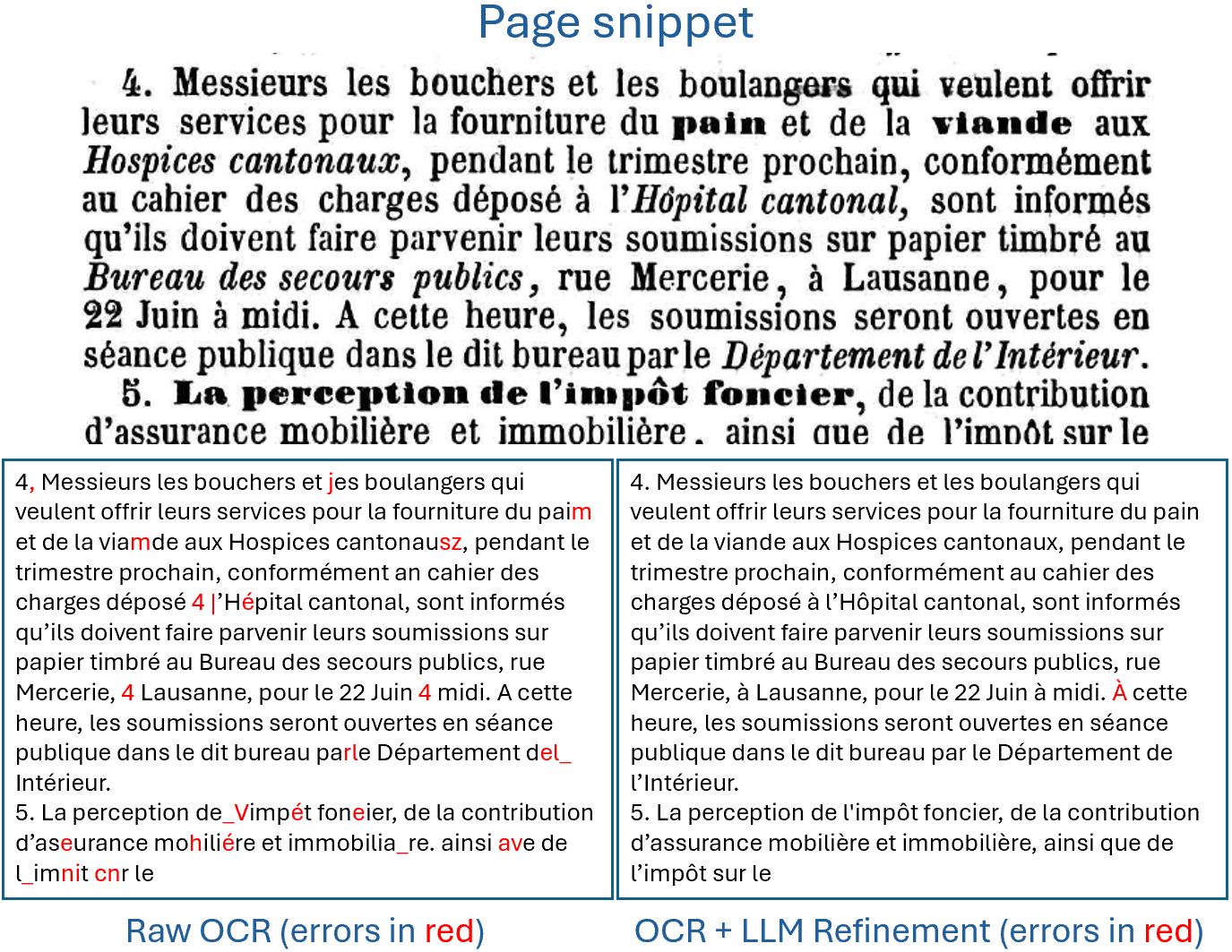}
    \caption{Example: Raw OCR (Tesseract) vs. raw OCR with LLM refinement (Gemini 3 Pro) for a page snippet.}
    \label{fig:ocr_example}
\end{figure}

Figure~\ref{fig:ocr_example} illustrates a sample case of text detection via OCR followed by LLM refinement. While the raw OCR output exhibits significant character-level errors, these are effectively mitigated in the LLM-refined version.

\subsection{Impact of Post-OCR LLM Corrections}

\noindent
\textbf{Results.}
We evaluate the improvement in text corrections for OCR systems, when post-processed via a variety of modern LLMs. These included widely adopted academic baselines \autocite{33418}, models developed by domain-specific companies, and solutions offered by major technology providers. For performance assessment, we used two standard evaluation metrics: word error rate (WER) and character error rate (CER). Both metrics quantify the accuracy of the predicted text relative to manually annotated ground truth data.

The comparative results of these models for the real-world historical newspaper dataset are presented in Tables~\ref{tab:CER_NEWSPAPERS} and~\ref{tab:WER_NEWSPAPERS}. We perform this experiment not on the full dataset, for which we do not have the ground-truth text, but on a random subset of 94 document snippets which we manually transcribed in order to have the ground-truth text.

OCR pipelines combined with LLM post-processing achieve a character error rate as low as $1.53\%$ and a word error rate as low as $4.93\%$. These error rates remain nontrivial; however, this is expected since the BCUL corpus comprises scanned historical documents set in old-style fonts with varying degradation patterns, posing significantly greater challenges. Therefore, these results should reflect realistic deployment conditions. What is important to note is that the integration of LLMs leads to drastic reductions in both WER and CER, compared to the raw OCR. The biggest improvements were seen with EasyOCR, while among the best-performing models in terms of average CER and WER reduction are Gemini 3.1 Pro ($44.52\%$) and Qwen3.7 Max ($60.95\%$) respectively, which also provide the most consistent improvements across OCR systems.

Interestingly, the results also highlight that not all LLMs are equally effective. While large, state-of-the-art models such as Gemini 3.1 Pro and Qwen3.7 Max deliver the strongest and most stable improvements, smaller or less instruction-tuned models, such as Qwen3 32B, often degrade performance, as seen most substantially with Apple Vision CER increasing from $3.66\%$ to $4.80\%$ and an average CER degradation of $-2.40\%$ across OCR engines. The reason is that smaller generative models are not able to follow the instructions for error correction, and instead of faithfully repairing OCR output, they tend to introduce additional text. This results in longer, noisier sequences that inflate both WER and CER, offsetting any benefits of correction. By contrast, larger LLMs with stronger instruction-following and editing capabilities are better able to constrain their outputs to faithful corrections, yielding consistent quality improvements in this more challenging, less curated setting.

To demonstrate the practical potential of the proposed system, we developed an immersive archive exploration interface deployed on an Apple Vision Pro device. The application enables users to query historical archives using natural language and receive answers grounded in retrieved documents.

The interface presents generated responses together with their supporting evidence, allowing users to inspect the original archival content and verify the provenance of each answer. In addition, contextual elements such as geographical references can be visualized to enhance exploration. This setup illustrates how combining LLM-based retrieval with interactive visualization can transform archival access into an exploratory and user-centric experience.

\begin{table*}[!h]
\centering
\setlength{\tabcolsep}{4pt}
\scriptsize
\caption{Historical newspapers CER (\%) using Raw OCR and OCR+LLM. Global best (lowest) is \textbf{bold}, 2nd-best is \underline{underlined}. Last row reports the average relative reduction (in \%) across OCR engines for each LLM. Last column reports the maximum relative reduction (in \%) with respect to the corresponding Raw OCR baseline for each OCR engine.}
\label{tab:CER_NEWSPAPERS}

\resizebox{\textwidth}{!}{%

\begin{tabular}{|l|r||r|r|r|r|r|r|r|r|r|r||r|}
\hline
 & \shortstack{Raw \\OCR} 
 & \shortstack{OCR + \\GPT-4o} 
 & \shortstack{OCR + \\GPT-5}
 & \shortstack{OCR + \\GPT-5.2}
 & \shortstack{OCR + \\Gemini 3\\Flash}
 & \shortstack{OCR + \\Gemini 3.1\\Pro}
 & \shortstack{OCR + \\Gemini 3.1\\Flash Lite}
 & \shortstack{OCR + \\Gemma 4}
 & \shortstack{OCR + \\DeepSeek\\V4}
 & \shortstack{OCR + \\Qwen3\\32B}
 & \shortstack{OCR + \\Qwen3.7\\Max}
 & \shortstack{Best $\Delta$ vs\\Raw OCR (\%)} \\
\hline

Tesseract
& 9.34 & 7.97 & 6.93 & 6.93 & \textbf{6.24} & \underline{6.34} & 6.71 & 6.57 & 7.44 & 9.09 & 6.48 & 33.20 \\

EasyOCR
& 10.88 & 5.09 & 4.21 & 4.24 & \underline{4.08} & \textbf{3.41} & 4.14 & 4.63 & 5.30 & 7.63 & 4.13 & 68.64 \\

Apple Vision OCR
& 3.66 & 4.11 & 3.21 & \underline{2.71} & 2.94 & \textbf{2.46} & 3.10 & 3.03 & 3.19 & 4.80 & 2.72 & 32.81 \\
\hline
Azure Document Intelligence
& 4.63 & 3.65 & 3.04 & 2.61 & 2.56 & \underline{2.35} & 2.54 & 2.86 & 2.68 & 4.42 & \textbf{2.23} & 51.82 \\
Google Document AI
& 2.90 & 3.03 & 1.93 & \underline{1.71} & 1.95 & 1.75 & 1.82 & 1.91 & 2.24 & 3.42 & \textbf{1.53} & 47.25 \\
\hline\hline
\shortstack{Average $\Delta$\\vs Raw OCR (\%)}
& -- & 14.44 & 33.45 & 39.50 & 38.57 & \textbf{44.52} & 37.60 & 35.37 & 29.90 & -2.40 & \underline{43.50} & -- \\
\hline
\end{tabular}%
}
\end{table*}

%%%%%%%%%%%%%%%%%%%%%%%%%%%%%%%%%%%%%%%%%%%%%%%%%%%%%%%%%

\begin{table*}[!h]
\centering
\setlength{\tabcolsep}{4pt}
\scriptsize
\caption{Historical newspapers WER (\%). Same layout and metrics as Table~\ref{tab:CER_NEWSPAPERS}.}
\label{tab:WER_NEWSPAPERS}

\resizebox{\textwidth}{!}{%

\begin{tabular}{|l|r||r|r|r|r|r|r|r|r|r|r||r|}
\hline
 & \shortstack{Raw \\OCR} 
 & \shortstack{OCR + \\GPT-4o} 
 & \shortstack{OCR + \\GPT-5}
 & \shortstack{OCR + \\GPT-5.2}
 & \shortstack{OCR + \\Gemini 3\\Flash}
 & \shortstack{OCR + \\Gemini 3.1\\Pro}
 & \shortstack{OCR + \\Gemini 3.1\\Flash Lite}
 & \shortstack{OCR + \\Gemma 4}
 & \shortstack{OCR + \\DeepSeek\\V4}
 & \shortstack{OCR + \\Qwen3\\32B}
 & \shortstack{OCR + \\Qwen3.7\\Max}
 & \shortstack{Best $\Delta$ vs\\Raw OCR (\%)} \\
\hline

Tesseract
& 25.86 & 16.75 & 14.99 & 14.95 & \textbf{13.52} & 14.65 & 14.53 & 14.75 & 16.40 & 18.59 & \underline{14.06} & 47.70 \\

EasyOCR
& 53.16 & 15.53 & 13.80 & 14.00 & 13.72 & \textbf{12.59} & \underline{13.55} & 15.28 & 17.37 & 21.34 & 13.66 & 76.31 \\

Apple Vision OCR
& 12.30 & 9.85 & 7.45 & 6.87 & 6.96 & \textbf{5.62} & 8.02 & 7.01 & 7.62 & 11.66 & \underline{6.56} & 54.31 \\
\hline
Azure Document Intelligence
& 29.12 & 9.55 & 7.93 & 7.87 & 7.43 & \underline{7.26} & 7.49 & 7.96 & 7.82 & 10.94 & \textbf{6.53} & 77.59 \\
Google Document AI
& 12.49 & 8.62 & 6.22 & \underline{5.56} & 5.96 & 6.26 & 6.15 & 6.15 & 6.53 & 9.66 & \textbf{4.93} & 60.52 \\
\hline\hline
\shortstack{Average $\Delta$\\vs Raw OCR (\%)}
& -- & 44.81 & 55.70 & 57.69 & 58.42 & \underline{59.78} & 55.64 & 56.13 & 52.57 & 35.65 & \textbf{60.95} & -- \\
\hline
\end{tabular}%
}
\end{table*}

\noindent

\subsection{Impact on Information Retrieval}

\medskip
\noindent
\textbf{RAG on Historical Archives.}
We evaluate our system with three core RAGAS metrics~\autocite{es2024ragas}: \emph{answer correctness}, \emph{context relevancy} and \emph{answer relevancy}, as well as NDCG@$k$. Let $q$ be the user query, $\mathcal{C}=\{c_i\}_{i=1}^m$ the retrieved contexts, $a$ the generated answer, and $g$ a ground-truth reference answer.

\medskip
\noindent
\underline{Correctness}\footnote{\label{note:correctness}\url{https://docs.ragas.io/en/v0.1.21/concepts/metrics/answer_correctness.html}}
captures semantic alignment with the reference answer. It evaluates how well $a$ agrees with the ground truth $g$ in both factual content and semantic similarity. RAGAS combines two components: (i) factual agreement $\text{fact}(a,g)\in[0,1]$ as judged by an LLM, and (ii) semantic similarity $\text{sim}(a,g)\in[0,1]$ based on embeddings. With default weights $w_f=0.75$ and $w_s=0.25$:
\[
\text{Correctness}(a,g) = w_f \cdot \text{fact}(a,g) + w_s \cdot \text{sim}(a,g).
\]
This balances factual fidelity with semantic closeness.

\medskip
\noindent
\underline{Relevancy}\footnote{\label{note:relevancy}\url{https://docs.ragas.io/en/v0.1.21/concepts/metrics/answer_relevance.html}}
To assess the alignment of the RAG pipeline, we define a general \textit{relevancy} metric based on the cosine similarity of sentence embeddings. Let $\phi(\cdot)$ represent a sentence embedding model and $\cos(\cdot,\cdot)$ denote the cosine similarity. The normalized relevance score is defined as
\[
\text{Rel}(x, y) = \text{norm}\big(\cos(\phi(x), \phi(y))\big).
\]

We apply this mathematical framework to two distinct stages of the generation process:

\begin{itemize}
    \item \textbf{Answer Relevancy:} Measures query alignment by calculating $\text{Rel}(q, a)$. It evaluates how well the generated answer $a$ responds to the user query $q$, focusing on topical focus rather than factual precision.
    \item \textbf{Context Relevancy:} Measures retrieval quality by calculating $\text{Rel}(q, c)$. It evaluates the extent to which the retrieved context $c$ is pertinent to the user query $q$.
\end{itemize}

A higher score indicates that $y$ is more directly on-topic with respect to the question, even if it is not factually perfect. Both metrics range from $[0,1]$, where a higher score indicates superior alignment.

\medskip
\noindent
\underline{NDCG@k} (normalized discounted cumulative gain) measures the quality of a ranking by comparing it against an ideal ordering of relevant items. It accounts for both the relevance of items and their positions in the list, where the discounted cumulative gain (DCG) at rank $k$ is defined as
\[
\text{DCG}_k = \sum_{i=1}^{k} \frac{rel_i}{\log_2(i + 1)};
\qquad
\text{NDCG}_k = \frac{\text{DCG}_k}{\text{IDCG}_k},
\]
where $rel_i$ is the graded relevance of the result at position $i$. To ensure the metric is comparable across queries, it is normalized by the ideal discounted cumulative gain (IDCG), which represents the DCG of the results sorted in descending order of relevance.

A higher score (closer to 1) indicates that the ranking more closely matches the ideal order, where all highly relevant items are positioned at the top of the list. All metrics are in the $[0,1]$ range, where higher is better.

\subsection{System response time}
As discussed in Section~3, the total system response time is largely driven by two components: the reranker and the text generation stage.

Many RAG systems rely on the BGE reranker, which is considered one of the strongest open-source rerankers. In our experiments this model achieves a very high NDCG value at the cost of very high latency, with mean reranking times over 2.5 seconds per query, making it impractical for latency-sensitive applications.

\begin{figure}[!t]
    \centering
    \includegraphics[width=0.99\linewidth]{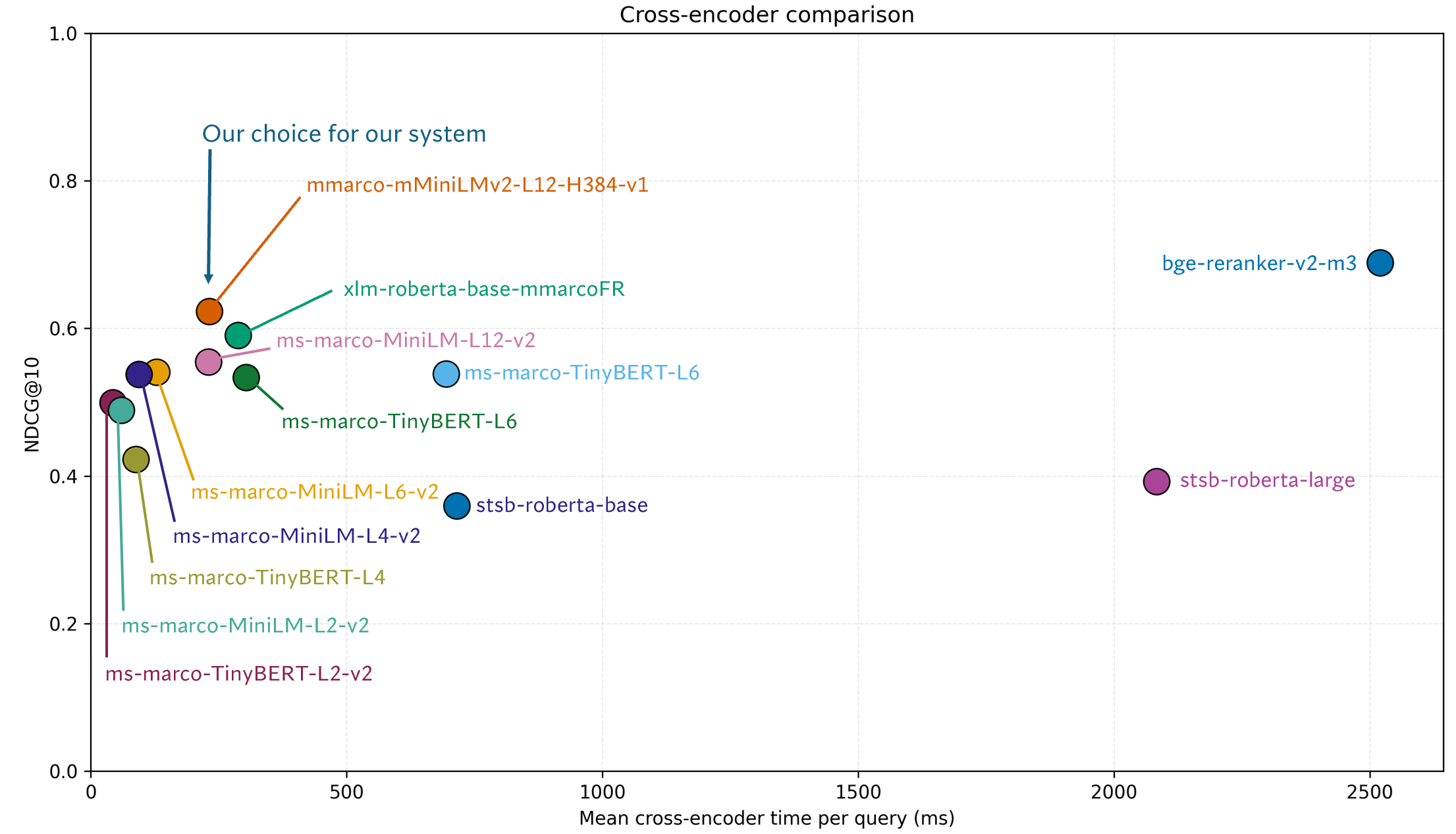}
    \caption{Reranking trade-off between quality (NDCG@10) and mean time (candidate documents $L=50$ used for reranking).}
    \label{fig:reranker_tradeoff}
\end{figure}

\FloatBarrier

To study the quality--latency trade-off, we evaluated several lighter MiniLM-based cross-encoders. Fig.~\ref{fig:reranker_tradeoff} reports NDCG@10 versus mean cross-encoder time computed over a large random set of queries. All models are compared under identical experimental conditions, taking 50 candidate documents as input and returning the top 10 results after reranking. While smaller models provide lower latency, they also show a noticeable drop in ranking quality. For our system deployment, \texttt{mmarco-mMiniLMv2-L12-H384-v1} offers a good compromise, achieving good reranking quality with a mean time slightly above 1 second. Although it mildly underperforms the BGE reranker in terms of ranking quality, it reduces latency significantly.

\begin{figure}[!t]
    \centering
    \includegraphics[width=0.99\linewidth]{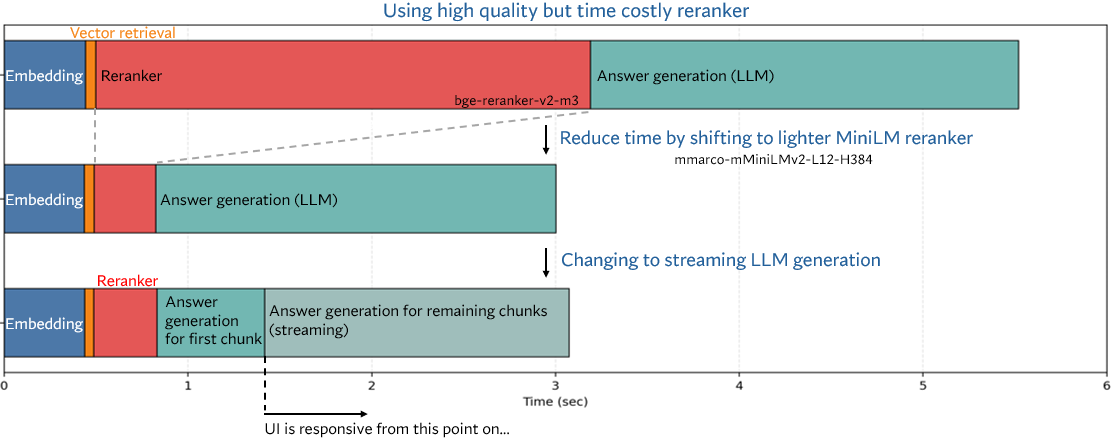}
    \caption{Deployed system, reducing system response time by changing the reranker and shifting to streaming LLM text generation (candidate documents $L=50$ for reranking).}
    \label{fig:time_reduction}
\end{figure}

\FloatBarrier

Figure \ref{fig:time_reduction} shows the time reduction when shifting to a lighter reranker for the deployed system where we use candidate documents $L=50$ for reranking, and by adopting a streaming LLM text generation. Now, the system receives responses after approximately 1.3 seconds.

In general, practitioners should carefully balance reranker quality and latency: large rerankers such as BGE may be suitable for offline or high-accuracy scenarios, whereas mid-size MiniLM models often provide a better trade-off for production RAG systems with strict response-time requirements. Switching to streaming LLM generation can also reduce the perceived response time of the interface.

Streaming generation does not significantly change total generation time but reduces time-to-first-token. This improves perceived responsiveness without affecting the underlying computation.

\subsection{Quality of answers and retrieved documents} 

Evaluating the quality of a new dataset is challenging because we have no ground-truth query--answer pairs. For this reason we adopt an LLM-as-a-judge evaluation using RAGAS metrics~\autocite{es2024ragas}, which estimate various aspects of system quality. Let $q$ denote the user query, $\mathcal{C}=\{c_i\}_{i=1}^m$ the retrieved contexts/documents, $a$ the generated answer, and $g$ a ground-truth reference answer. RAGAS uses GPT-4o-mini as the evaluation model.

We use the following \textbf{metrics}, all in $[0,1]$ (higher is better): \underline{Answer} \underline{Correctness}\textsuperscript{\ref{note:correctness}} combines LLM-judged factual agreement and embedding-based semantic similarity between the generated answer $a$ and ground truth $g$: $\text{Correctness}(a,g)=0.75\cdot\text{fact}(a,g)+0.25\cdot\text{sim}(a,g)$. \underline{Answer and Context} \underline{ Relevancy}\textsuperscript{\ref{note:relevancy}} measure, respectively, how well $a$ addresses the query $q$ and how pertinent the retrieved context $c$ is to $q$, both computed as normalized cosine similarities between sentence embeddings. \underline{NDCG@$k$} (Normalized Discounted Cumulative Gain) evaluates ranking quality by comparing the retrieved ranking against the ideal ordering of relevant items.

\medskip
\noindent
\section{Results}
Building on our previous findings, this experiment retained only some of the top-performing OCR and LLM models for a practical retrieval deployment. For raw text extraction, we use Apple Vision OCR, and use Gemini 3 Flash to sanitize transcription errors. In archival search systems, response latency and operational cost are critical deployment constraints, and Gemini 3 flash provided one of the most favorable trade-offs among correction quality, inference latency, rate limits, and cost efficiency. The resulting text was embedded using Google's embeddings 001, with OpenAI's GPT-4o-mini generating the final text from the retrieved documents for its reliability and consistent results.

\vspace{-\baselineskip}
\begin{table*}[h]
\centering
\footnotesize
\setlength{\tabcolsep}{6pt}
\renewcommand{\arraystretch}{1.1}
\resizebox{\textwidth}{!}{%
\begin{tabular}{clccc}
\toprule
\textbf{$k$} & \textbf{Method} & \textbf{Context Relevance} & \textbf{Answer Correctness} & \textbf{NDCG@k} \\
\midrule
5 & BM25s &  69.27$\pm$3.73\% & 52.22$\pm$3.09\% & 65.14$\pm$4.22\% \\
 & LLM(BM25s)  & 67.97$\pm$3.70\% & 53.12$\pm$3.21\% & 64.78$\pm$4.22\% \\
 & Semantic  & 67.32$\pm$3.54\% & 51.69$\pm$3.32\% & 63.10$\pm$4.33\% \\
 & LLM(Semantic)  & 69.08$\pm$3.71\% & 50.66$\pm$3.24\% & 62.37$\pm$4.27\% \\
 & Semantic + BM25s &  73.83$\pm$3.53\%\textsuperscript{***} & 60.22$\pm$2.94\%\textsuperscript{***} & 75.42$\pm$3.82\%\textsuperscript{***} \\
 & LLM(Semantic + BM25s) & 75.20$\pm$3.45\% & 60.69$\pm$3.02\% & 71.50$\pm$3.72\% \\
 & Semantic + BM25s + reranker &  83.92$\pm$2.65\%\textsuperscript{***} & 64.71$\pm$2.63\%\textsuperscript{***} & 85.73$\pm$3.01\%\textsuperscript{***} \\
 & \textbf{Ours - LLM(Semantic + BM25s) + reranker}  & \textbf{86.72$\pm$2.59\%}\textsuperscript{***} & \textbf{67.40$\pm$2.62\%}\textsuperscript{***} & \textbf{87.79$\pm$2.90\%}\textsuperscript{***} \\
\hline
10 & BM25s &  70.51$\pm$3.65\% & 56.15$\pm$2.95\% & 65.99$\pm$3.86\% \\
 & LLM(BM25s)  & 69.60$\pm$3.69\% & 57.45$\pm$2.98\%\textsuperscript{*} & 65.51$\pm$3.86\% \\
 & Semantic &  71.29$\pm$3.51\% & 54.73$\pm$3.13\% & 63.69$\pm$4.06\% \\
 & LLM(Semantic) & 73.18$\pm$3.55\%\textsuperscript{*} & 54.74$\pm$2.99\% & 62.93$\pm$3.95\% \\
 & Semantic + BM25s  & 75.39$\pm$3.40\% & 61.43$\pm$2.74\%\textsuperscript{***} & 74.35$\pm$3.59\%\textsuperscript{***} \\
 & LLM(Semantic + BM25s)  & 77.99$\pm$3.39\%\textsuperscript{**} & 62.62$\pm$2.77\% & 69.71$\pm$3.38\% \\
 & Semantic + BM25s + reranker & 85.03$\pm$2.57\%\textsuperscript{***} & 63.60$\pm$2.52\% & 84.94$\pm$2.84\%\textsuperscript{***} \\
 & \textbf{Ours - LLM(Semantic + BM25s) + reranker} &  \textbf{87.63$\pm$2.57\%}\textsuperscript{***} & \textbf{65.92$\pm$2.57\%}\textsuperscript{**} & \textbf{87.05$\pm$2.77\%}\textsuperscript{***} \\
\hline
25 & BM25s  & 69.99$\pm$3.86\% & 59.25$\pm$2.77\% & 65.77$\pm$3.55\% \\
 & LLM(BM25s)  & 67.58$\pm$3.89\% & 59.42$\pm$2.84\% & 65.36$\pm$3.57\% \\
 & Semantic  & 72.14$\pm$3.53\%\textsuperscript{**} & 58.33$\pm$2.93\% & 61.91$\pm$3.75\% \\
 & LLM(Semantic)  & 75.85$\pm$3.45\%\textsuperscript{***} & 58.90$\pm$2.78\% & 62.06$\pm$3.68\% \\
 & Semantic + BM25s  & 77.35$\pm$3.28\% & 64.78$\pm$2.48\%\textsuperscript{***} & 72.91$\pm$3.19\%\textsuperscript{***} \\
 & LLM(Semantic + BM25s)  & 81.07$\pm$4.56\%\textsuperscript{**} & 63.81$\pm$4.31\% & 66.93$\pm$2.86\% \\
 & Semantic + BM25s + reranker  & 84.51$\pm$2.65\% & 63.91$\pm$2.56\%\textsuperscript{**} & 82.02$\pm$2.75\%\textsuperscript{***} \\
 & \textbf{Ours - LLM(Semantic + BM25s) + reranker}  & \textbf{87.17$\pm$2.57\%}\textsuperscript{***} & \textbf{65.96$\pm$2.56\%}\textsuperscript{**} & \textbf{84.79$\pm$2.66\%}\textsuperscript{***} \\
\bottomrule
\end{tabular}
}
\caption{Ablation and performance comparison across retrieval depths ($k \in \{5,10, 25\}$) with and without reranking. LLM($\cdot$) denotes the same retrieval configuration applied to the LLM-corrected OCR view of the corpus. Dense retrieval uses text-embedding-2-preview, and answer generation is performed using GPT-4o-mini. The $\pm$ limits denote the 95\% confidence interval. Superscripts $^{*}$ , $^{**}$, and $^{***}$ indicate statistically significant improvement over the second-best method at levels $p < 0.1$, $p < 0.05$, and $p < 0.01$ respectively, according to a paired t-test.}
\label{tab:IRexperimentresults}
\end{table*}

Table~\ref{tab:IRexperimentresults} reports the performance across 384 queries for different retrieval depths ($k \in \{5,10, 25\}$), comparing lexical retrieval (BM25s), standard dense retrieval (Semantic), dense retrieval on LLM-corrected OCR (LLM(Semantic)), and a cross-encoder reranking pipeline (Reranker). The LLM-corrected corpus uses Gemini 3 Flash for OCR sanitization.

The data demonstrates that BM25s provides a competitive lexical baseline, but the semantic and reranking methods provide substantial relative improvements; for instance, at $k=10$, our full pipeline outperforms BM25s in NDCG by a relative $31.91\%$ ($65.99\% \rightarrow 87.05\%$) and Context Relevance by a relative $24.28\%$ ($70.51\% \rightarrow 87.63\%$).

In order to further enhance our reranking strategy, metadata such as publication date, location or entities relating to a given article were provided, allowing efficient dismissal of records not conforming to the criteria set by the users prompt.

When analyzing the dense retrieval strategies, applying LLM-corrected OCR (LLM(Semantic)) yields targeted improvements over the standard Semantic baseline. For example, at $k=10$, the LLM-corrected pipeline achieves a $2.65\%$ relative improvement in Context Relevance ($71.29\% \rightarrow 73.18\%$). The effect is more pronounced when combining dense and lexical retrieval: at $k=10$, LLM(Semantic + BM25s) improves Context Relevance by $3.45\%$ over its uncorrected counterpart ($75.39\% \rightarrow 77.99\%$) and Answer Correctness by $1.94\%$ ($61.43\% \rightarrow 62.62\%$). While the uncorrected Semantic baseline remains competitive, the LLM-corrected version consistently reduces noise, leading to more query-aligned responses.

In summary, while LLM correction of OCR text provides specific stabilizing benefits for answer accuracy, the addition of a reranker is the definitive driver of overall retrieval quality. The reranker pipeline consistently dominates complex metrics, delivering a $24.65\%$ relative increase in Context Relevance over the Semantic baseline at $k=5$ ($67.32\% \rightarrow 83.92\%$) and culminating in statistically significant, peak performances for NDCG ($87.79\%$ at $k=5$) and Answer Correctness ($67.40\%$ at $k=5$). Overall, these results demonstrate that cleaning OCR text with LLMs improves the factual grounding of responses (Correctness), while the reranker ensures those responses are highly query-aligned (Context Relevance).

\begin{figure}[!t]
    \centering
    \includegraphics[width=.99\linewidth]{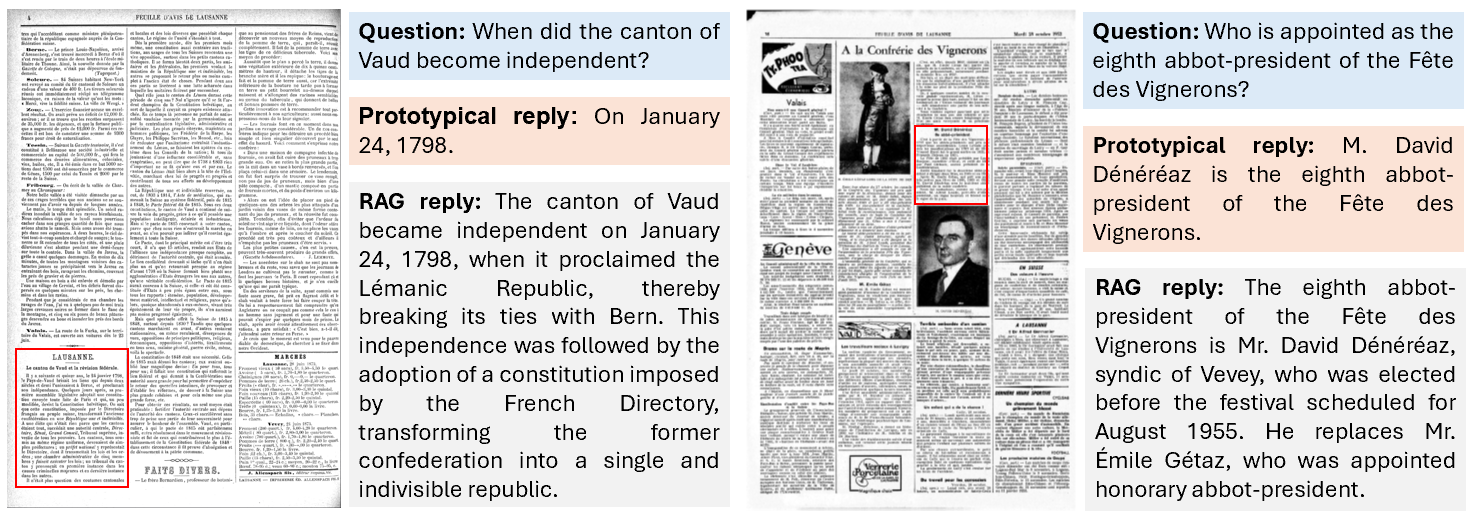}
    \caption{Examples of two queries, the prototypical and the actual reply by the RAG system. We also show indicative documents retrieved by the search process.}
    \label{fig:rag_illustrations}
\end{figure}

\subsection{Qualitative Results}

Figure~\ref{fig:rag_illustrations} presents two representative examples of queries paired with retrieved documents and system replies, highlighting the qualitative impact of retrieval-augmented generation.

In Query~1, which asks when the canton of Vaud became independent, the RAG system retrieves a historical newspaper excerpt describing the proclamation of independence on January 24, 1798. Compared to a prototypical LLM reply, which includes additional contextual details but may introduce temporal ambiguity or unnecessary elaboration, the RAG-based answer is more concise and directly grounded in the retrieved source, accurately identifying the key date and event.

Query~2 illustrates a multi-document factual retrieval scenario concerning the Fêtes des Vignerons between 1900 and 1970. The system aggregates information from several newspaper issues to correctly identify the relevant years (1905, 1927, and 1955). In contrast, the prototypical reply presents fragmented or less structured information, whereas the RAG system produces a coherent and consolidated answer grounded in archival evidence.

Taken together, these examples illustrate how RAG systems improve factual accuracy and answer reliability by anchoring responses in retrieved historical documents. While standalone LLMs may generate fluent but sometimes diffuse or partially inconsistent answers, retrieval-augmented generation enables more precise, verifiable, and contextually grounded outputs, particularly when dealing with noisy OCR data from archival sources.

\subsection{Application-Level Demonstration}

To illustrate the practical applicability of the proposed pipeline, we implemented a prototype interface for interactive exploration of archival data on an Apple Vision Pro device. The system allows users to issue natural-language queries and receive generated answers together with the supporting archival evidence, while benefiting from a more immersive interaction model in augmented reality.

\begin{figure*}
    \centering
    \includegraphics[width=0.99\linewidth]{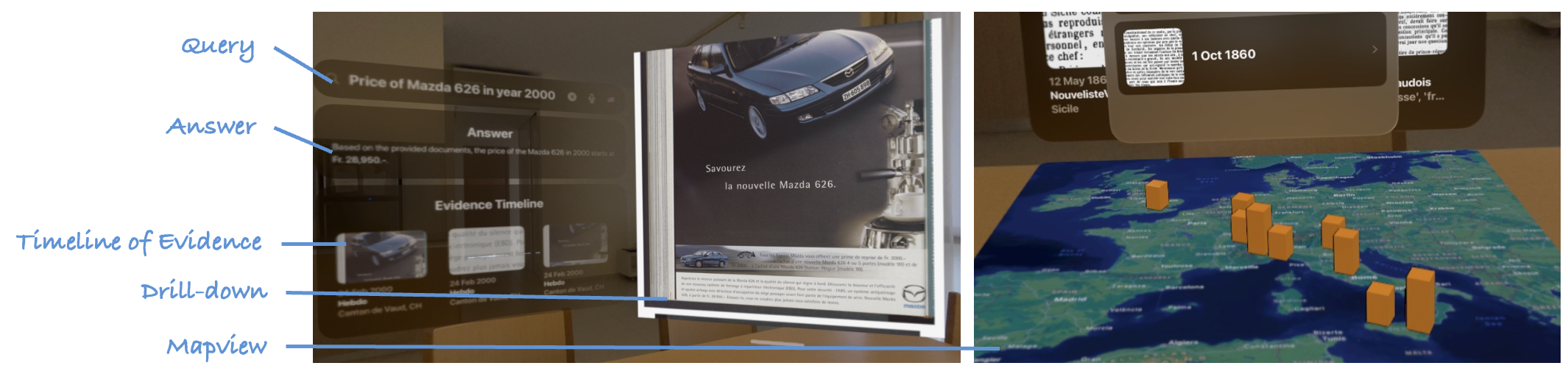}
    \caption{Prototypical interface using the Vision Pro augmented reality device}
    \label{fig:visionpro}
\end{figure*}

The interface (see Figure \ref{fig:visionpro}) exposes the retrieved documents used during generation, enabling users to verify the provenance of the responses. It also supports further navigation into the original archival material, facilitating in-depth inspection of the source documents. In addition, the spatial interface enables a geographical mapping of the locations mentioned in the documents, thus allowing to explore the locations involved in a particular query. This prototype illustrates how the proposed pipeline can be integrated into user-facing systems for exploratory access to historical archives. A video of the interface is available here: 
\href{https://youtu.be/jDUzbM4BmF0}{Apple Vision video application}.

\section{Discussion and Implications}
\label{sec:discussion}
\subsection{Discussion of Results}
The results of our evaluation highlight the interconnected nature of data ingestion and query execution in historical archival systems. First, regarding OCR sanitization, our findings help resolve conflicting reports in recent literature by demonstrating that the success of LLM-based OCR correction is highly dependent on model scale and instruction-following capabilities. While smaller models (e.g., Qwen3 32B) tend to over-correct and degrade text fidelity, frontier models (e.g., Gemini 3.1 Pro, Qwen3.7 Max) yield substantial reductions in both Character and Word Error Rates (up to $44.52\%$ average CER reduction and $60.95\%$ average WER reduction).
Crucially, our experiments validate that these ingestion-level improvements directly propagate to query-time performance. However, our ablation study in Table~\ref{tab:IRexperimentresults} reveals a nuanced hierarchy of impact: while LLM-based OCR sanitization primarily drives improvements in the factual grounding of generated answers (\textit{Answer Correctness}), it is the integration of a cross-encoder reranking pipeline that acts as the primary catalyst for overall retrieval quality (\textit{Context Relevancy} and \textit{NDCG}).
\subsection{Evaluation Implications}
This research diverges from and extends existing work in several theoretical dimensions. Historically, the document analysis community has treated OCR error correction as an isolated task evaluated purely on lexical fidelity (CER/WER). Conversely, the information retrieval community often evaluates RAG architectures assuming clean, contemporary text corpora. 
Our work bridges this theoretical gap by proposing an end-to-end evaluation paradigm. We demonstrate that for historical archives, OCR quality should not be evaluated in a vacuum, but rather by its downstream impact on semantic retrieval and generative question answering. Furthermore, by formalizing the trade-offs between dense retrieval pool size, reranking complexity, and generation latency, we provide a new framework for modeling the computational bottlenecks inherent in applying RAG to massive, noisy historical datasets.
\subsection{Practical Implications}
From a practical standpoint, our findings offer a blueprint for modernizing digital libraries. For archivists and platform maintainers, our results demonstrate that perfect, human-curated OCR may not always be required to obtain useful semantic search results, especially when retrieval is combined with reranking and grounded generation. By employing an architecture that pairs automated LLM-sanitization with robust reranking, institutions can unlock access to noisy heritage collections that were previously undiscoverable via rigid keyword search.
Furthermore, our latency-aware design, which uses lightweight cross-encoders and streaming token generation, proves that advanced RAG pipelines can be deployed under strict response-time constraints. As demonstrated by our immersive Apple Vision Pro interface, this enables digital libraries to transition from static, 2D repositories into highly interactive, real-time exploratory environments, fundamentally changing how historians, journalists, and the public engage with cultural heritage data.

\section{Conclusion}

We showed that large language models can significantly improve text quality and semantic access in large-scale digital archives. We introduced an end-to-end document processing pipeline that integrates LLMs into the archival workflow, enabling post-OCR corrections and enhanced retrieval-augmented semantic querying. Experiments on a real-world archival dataset of Swiss newspapers demonstrate improved transcription quality and downstream document retrieval and question answering, increasing the correctness and relevance of generated responses. 

Our evaluation indicates that for standalone archival digitisation, LLM-based post-OCR refinement provides a critical boost to text fidelity and readability. However, in scenarios where the primary goal is downstream retrieval-augmented generation (RAG), the inherent reasoning capabilities of the LLM can often compensate for original OCR noise, making preemptive text correction less essential. Consequently, while refinement is vital for high-quality archival preservation, its necessity diminishes when the objective is restricted to broad semantic querying within a RAG framework.
From a document analysis perspective, our findings highlight LLMs as practical components for text correction and semantic access, supporting future intelligent archival pipelines.

\section*{Acknowledgments}
This work was supported by the Swiss National Science Foundation (SNSF), Switzerland, under Grant No. 237991.

\section*{AI-Generated Content Acknowledgement}
Generative AI tools were used for grammar refinement, clarity improvement, and English language checking throughout the manuscript. All research ideas and methods are entirely the authors' own.

\section*{Data Availability Statement}
The archival newspaper data used in this study were provided by the Cantonal and University Library of Lausanne through the Scriptorium portal. Due to rights and data-owner restrictions, the full dataset cannot be publicly redistributed at this stage. The authors plan to release a subset of the dataset upon clearance from the data owner. The code used for the experiments is available in the \href{https://github.com/Stergios-Konstantinidis/Improving-Access-to-Vast-Historical-Archives-with-Large-Language-Models}{code repository}.

\section*{Conflict of Interest Statement}
The authors declare no conflicts of interest.

\theendnotes

\printbibliography

@article{transkribus2022,
  title     = {{Understanding the application of handwritten text recognition technology in heritage contexts: a systematic review of Transkribus in published research}},
  author    = {Nockels, Joe and Gooding, Paul and Ames, Sarah and Terras, Melissa},
  journal   = {Archival Science},
  pages     = {367-392},
  year      = {2022},
  publisher = {Springer},
  doi={10.1007/s10502-022-09397-0}
}

@inproceedings{lee2020newspaper,
author = {Lee, Benjamin Charles Germain and Mears, Jaime and Jakeway, Eileen and Ferriter, Meghan and Adams, Chris and Yarasavage, Nathan and Thomas, Deborah and Zwaard, Kate and Weld, Daniel S.},
title = {The Newspaper Navigator Dataset: Extracting Headlines and Visual Content from 16 Million Historic Newspaper Pages in Chronicling America},
year = {2020},
isbn = {9781450368599},
publisher = {Association for Computing Machinery},
doi = {10.1145/3340531.3412767},
booktitle = {Proceedings of the 29th ACM International Conference on Information \& Knowledge Management},
pages = {3055–3062},
numpages = {8}
}

@article{johnson2019billion,
  author  = {Johnson, Jeff and Douze, Matthijs and Jégou, Hervé},
  journal = {IEEE Transactions on Big Data},
  title   = {{Billion-scale similarity search with GPUs}},
  year    = {2021},
  volume  = {7},
  number  = {03},
  pages   = {535-547},
  url     = {https://doi.ieeecomputersociety.org/10.1109/TBDATA.2019.2921572}
}

@article{banerjee2024making,
  title={Making History Readable},
  author={Bipasha Banerjee and Jennifer Goyne and William A. Ingram},
  journal={IEEE International Conference on Big Data},
  year={2024},
  pages={8620-8622},
  url={https://api.semanticscholar.org/CorpusID:274281491}
}

@article{jaillant2022applying,
    author = {Jaillant, Lise and Rees, Arran},
    title = {{Applying AI to digital archives: trust, collaboration and shared professional ethics}},
    journal = {Digital Scholarship in the Humanities},
    volume = {38},
    number = {2},
    pages = {571-585},
    year = {2023},
    month = {06},
    doi   = {10.1093/llc/fqac073}
}

@inproceedings{groppe2025automated,
author = {Zhang, Hao},
title = {Towards a Digital Archivist: Applications of LLMs in Automated Web Archive Description},
year = {2025},
url = {https://doi.org/10.1007/978-981-95-4861-3_40},
doi = {10.1007/978-981-95-4861-3_40},
booktitle = {Intelligence and Equity: Shaping the Future of Knowledge: 27th International Conference on Asian Digital Libraries, ICADL},
pages = {474–483},
location = {Metro Manila, Philippines}
}

@Article{ehrmanncomputational,
  author    =	{Ehrmann, Maud and D\"{u}ring, Marten and Neudecker, Clemens and Doucet, Antoine},
  title     =	{{Computational Approaches to Digitised Historical Newspapers}},
  pages     =	{112--179},
  journal   =	{Dagstuhl Reports},
  ISSN      =	{2192-5283},
  year      =	{2023},
  volume    =	{12},
  number    =	{7},
  publisher =	{Schloss Dagstuhl -- Leibniz-Zentrum f{\"u}r Informatik},
  address   =	{Dagstuhl, Germany},
  URL       =	{https://drops.dagstuhl.de/entities/document/10.4230/DagRep.12.7.112},
  URN       =	{urn:nbn:de:0030-drops-176141},
  doi       =	{10.4230/DagRep.12.7.112},
}

@inproceedings{boros_ehrmann2024,
    title = "Post-Correction of Historical Text Transcripts with Large Language Models: An Exploratory Study",
    author = "Boros, Emanuela  and
      Ehrmann, Maud  and
      Romanello, Matteo  and
      Najem-Meyer, Sven  and
      Kaplan, Fr{\'e}d{\'e}ric",
    booktitle = "Proceedings of the 8th Joint SIGHUM Workshop on Computational Linguistics for Cultural Heritage, Social Sciences, Humanities and Literature",
    month = mar,
    year = "2024",
    url = "https://aclanthology.org/2024.latechclfl-1.14/",
    doi = "10.18653/v1/2024.latechclfl-1.14",
    pages = "133--159",
}

@techreport{docling,
  author = {Deep Search Team},
  month = {8},
  title = {Docling Technical Report},
  url = {https://arxiv.org/abs/2408.09869},
  eprint = {2408.09869},
  doi = {10.48550/arXiv.2408.09869},
  version = {1.0.0},
  year = {2024},
  institution = {IBM}
}

@inproceedings{es2024ragas,
   author = {Shahul Es and Jithin James and Luis Espinosa-Anke and Steven Schockaert},
   doi = {10.18653/V1/2024.EACL-DEMO.16},
   isbn = {9798891760912},
   journal = {EACL 2024 - 18th Conference of the European Chapter of the Association for Computational Linguistics, Proceedings of System Demonstrations},
   pages = {150-158},
   publisher = {Association for Computational Linguistics},
   title = {{RAGAs: Automated evaluation of retrieval augmented generation}},
   url = {https://aclanthology.org/2024.eacl-demo.16/},
   year = {2024}
}

@misc{zhao2023survey,
      title={A Survey of Large Language Models}, 
      author={Wayne Xin Zhao and Kun Zhou and Junyi Li and Tianyi Tang and Xiaolei Wang and Yupeng Hou and Yingqian Min and Beichen Zhang and Junjie Zhang and Zican Dong and Yifan Du and Chen Yang and Yushuo Chen and Zhipeng Chen and Jinhao Jiang and Ruiyang Ren and Yifan Li and Xinyu Tang and Zikang Liu and Peiyu Liu and Jian-Yun Nie and Ji-Rong Wen},
      year={2026},
      eprint={2303.18223},
      archivePrefix={arXiv},
      primaryClass={cs.CL},
      url={https://arxiv.org/abs/2303.18223}, 
}

@inproceedings{fan2024survey,
author = {Fan, Wenqi and Ding, Yujuan and Ning, Liangbo and Wang, Shijie and Li, Hengyun and Yin, Dawei and Chua, Tat-Seng and Li, Qing},
title = {A Survey on RAG Meeting LLMs: Towards Retrieval-Augmented Large Language Models},
year = {2024},
isbn = {9798400704901},
publisher = {Association for Computing Machinery},
url = {https://doi.org/10.1145/3637528.3671470},
doi = {10.1145/3637528.3671470},
booktitle = {Proceedings of the ACM SIGKDD International Conference on Knowledge Discovery and Data Mining},
pages = {6491–6501},
volume = {24},
numpages = {11},

}

@article{barman2021,
author = {Barman, Raphaël and Ehrmann, Maud and Clematide, Simon and Oliveira, Sofia and Kaplan, Frederic},
year = {2021},
month = {01},
pages = {},
title = {Combining Visual and Textual Features for Semantic Segmentation of Historical Newspapers},
volume = {HistoInformatics},
journal = {Journal of Data Mining \& Digital Humanities},
doi = {10.46298/jdmdh.6107}
}

@article{chowdhery2023,
author = {Chowdhery, Aakanksha and Narang, Sharan and Devlin, Jacob and Bosma, Maarten and Mishra, Gaurav and Roberts, Adam and Barham, Paul and Chung, Hyung Won and Sutton, Charles and Gehrmann, Sebastian and Schuh, Parker and Shi, Kensen and Tsvyashchenko, Sashank and Maynez, Joshua and Rao, Abhishek and Barnes, Parker and Tay, Yi and Shazeer, Noam and Prabhakaran, Vinodkumar and Reif, Emily and Du, Nan and Hutchinson, Ben and Pope, Reiner and Bradbury, James and Austin, Jacob and Isard, Michael and Gur-Ari, Guy and Yin, Pengcheng and Duke, Toju and Levskaya, Anselm and Ghemawat, Sanjay and Dev, Sunipa and Michalewski, Henryk and Garcia, Xavier and Misra, Vedant and Robinson, Kevin and Fedus, Liam and Zhou, Denny and Ippolito, Daphne and Luan, David and Lim, Hyeontaek and Zoph, Barret and Spiridonov, Alexander and Sepassi, Ryan and Dohan, David and Agrawal, Shivani and Omernick, Mark and Dai, Andrew M. and Pillai, Thanumalayan Sankaranarayana and Pellat, Marie and Lewkowycz, Aitor and Moreira, Erica and Child, Rewon and Polozov, Oleksandr and Lee, Katherine and Zhou, Zongwei and Wang, Xuezhi and Saeta, Brennan and Diaz, Mark and Firat, Orhan and Catasta, Michele and Wei, Jason and Meier-Hellstern, Kathy and Eck, Douglas and Dean, Jeff and Petrov, Slav and Fiedel, Noah},
   journal = {Journal of Machine Learning Research},
   pages = {1-113},
   title = {{PaLM: Scaling language modeling with pathways}},
   volume = {24},
   url = {https://jmlr.org/papers/v24/22-1144.html},
   year = {2023}
}

@misc{spina2023archival,
      title={Artificial Intelligence in archival and historical scholarship workflow: HTS and ChatGPT}, 
      author={Salvatore Spina},
      year={2023},
      eprint={2308.02044},
      archivePrefix={arXiv},
      primaryClass={cs.DL},
      url={https://arxiv.org/abs/2308.02044}, 
}

@inproceedings{thomas2024,
    title = "Leveraging {LLM}s for Post-{OCR} Correction of Historical Newspapers",
    author = "Thomas, Alan and Gaizauskas, Robert and Lu, Haiping",
    booktitle = "Proceedings of the Third Workshop on Language Technologies for Historical and Ancient Languages (LT4HALA)",
    url = "https://aclanthology.org/2024.lt4hala-1.14/",
    pages = "116-121",
    year = {2024}
}

@inproceedings{do2025,
author = {Do, Thao and Tran, Dinh Phu and Vo, An and Kim, Daeyoung},
title = {{Reference-based post-OCR processing with LLM for precise diacritic text in historical document recognition}},
year = {2025},
isbn = {157735897X},
url = {https://doi.org/10.1609/aaai.v39i27.35012},
pages = {27951-27959},
doi = {10.1609/aaai.v39i27.35012},
volume = {39},
publisher = {Association for the Advancement of Artificial Intelligence},
journal = {Proceedings of the AAAI Conference on Artificial Intelligence},
}

@inproceedings{manrique2024historical,
    title = "Historical Ink: 19th Century {L}atin {A}merican {S}panish Newspaper Corpus with {LLM} {OCR} Correction",
    author = "Manrique-Gomez, Laura  and
      Montes, Tony  and
      Rodriguez Herrera, Arturo  and
      Manrique, Ruben",
    booktitle = "Proceedings of the 4th International Conference on Natural Language Processing for Digital Humanities",
    month = nov,
    year = "2024",
    url = "https://aclanthology.org/2024.nlp4dh-1.13/",
    doi = "10.18653/v1/2024.nlp4dh-1.13",
    pages = "132--139"
}

@inproceedings{shen_layoutparser:_2021,
author="Shen, Zejiang
and Zhang, Ruochen
and Dell, Melissa
and Lee, Benjamin Charles Germain
and Carlson, Jacob
and Li, Weining",
title="LayoutParser: A Unified Toolkit for Deep Learning Based Document Image Analysis",
booktitle="Document Analysis and Recognition -- ICDAR",
year="2021",
pages="131--146",
isbn={"978-3-030-86549-8"},
doi={10.1007/978-3-030-86549-8_9}
}

@inproceedings{doclaynet2022,
author = {Pfitzmann, Birgit and Auer, Christoph and Dolfi, Michele and Nassar, Ahmed S. and Staar, Peter},
title = {DocLayNet: A Large Human-Annotated Dataset for Document-Layout Segmentation},
year = {2022},
isbn = {9781450393850},
url = {https://doi.org/10.1145/3534678.3539043},
doi = {10.1145/3534678.3539043},
booktitle = {Proceedings of the 28th ACM SIGKDD Conference on Knowledge Discovery and Data Mining},
pages = {3743–3751},
numpages = {9},
}

@inproceedings{33418,
author = {Smith, R.},
title = {An Overview of the Tesseract OCR Engine},
year = {2007},
isbn = {0769528228},
booktitle = {Proceedings of the Ninth International Conference on Document Analysis and Recognition},
pages = {629–633},
numpages = {5},
series = {ICDAR},
}

@misc{sahoo_systematic_2025,
      title={A Systematic Survey of Prompt Engineering in Large Language Models: Techniques and Applications}, 
      author={Pranab Sahoo and Ayush Kumar Singh and Sriparna Saha and Vinija Jain and Samrat Mondal and Aman Chadha},
      year={2025},
      eprint={2402.07927},
      archivePrefix={arXiv},
      primaryClass={cs.AI},
      url={https://arxiv.org/abs/2402.07927}, 
}

@misc{gemma_2025,
  author  = {{Gemma Team}},
  title   = {Gemma 3 Technical Report},
  year    = {2025},
  url     = {https://doi.org/10.48550/arXiv.2503.19786},
  url     ={https://ai.google.dev/gemma/docs/core/model_card_3}
}

@misc{comanici_gemini_2025,
  author = {{Gemini Team}},
  title  = {Gemini 2.5 Pro Model Card},
  year   = {2025},
  url    = {https://storage.googleapis.com/deepmind-media/Model-Cards/Gemini-2-5-Pro-Model-Card.pdf}
}

@misc{gemini3pro,
  author       = {{Gemini Team}},
  title        = {Gemini 3 Pro Model Card},
  year         = {2025},
  url          = {https://storage.googleapis.com/deepmind-media/Model-Cards/Gemini-3-Pro-Model-Card.pdf}
}

@misc{gemini3flash,
  author       = {{Gemini Team}},
  title        = {Gemini 3 Flash Model Card},
  year         = {2025},
  url          ={https://storage.googleapis.com/deepmind-media/Model-Cards/Gemini-3-Flash-Model-Card.pdf}
}

@misc{easyocr,
  author = {{JaidedAI}},
  title  = {{EasyOCR: Ready-to-use OCR}},
  year   = {2020},
  url     = {https://github.com/JaidedAI/EasyOCR}
}

@misc{qwen3technicalreport,
  title  = {{Qwen3 Technical report}},
  author = {{Qwen Team}},
  year   = {2025},
  doi    = {10.48550/arXiv.2505.09388}
}

@misc{deepseekai2025deepseekr1incentivizingreasoningcapability,
  title  = {{DeepSeek-R1 Technical report}},
  author = {{DeepSeek Team}},
  year   = {2025},
  url    = {https://github.com/deepseek-ai/DeepSeek-R1/blob/main/DeepSeek_R1.pdf}
}

@techreport{openai_gpt-4_2024,
  title  = {{GPT-4 Technical report}},
  author = {{OpenAI}},
  year   = {2024},
  doi    = {10.48550/arXiv.2303.08774},
  url = {https://arxiv.org/pdf/2303.08774}
}

@techreport{openai_gpt5_2025,
  title  = {{GPT-5 System card}},
  author = {{OpenAI}},
  year   = {2025},
  doi    = {10.48550/arXiv.2601.03267},
  url    = {https://arxiv.org/pdf/2601.03267}
}

@techreport{openai_gpt52_2025,
  author = {{OpenAI}},
  title  = {{GPT-5.2 System card}},
  year   = {2025},
  url = {https://cdn.openai.com/pdf/3a4153c8-c748-4b71-8e31-aecbde944f8d/oai_5_2_system-card.pdf}
}

@misc{zhang2025ocr,
  title     = {{OCR hinders RAG: Evaluating the cascading impact of ocr on retrieval-augmented generation}},
  author    = {Zhang, Junyuan and Zhang, Qintong and Wang, Bin and Ouyang, Linke and Wen, Zichen and Li, Ying and Chow, Ka-Ho and He, Conghui and Zhang, Wentao},
  booktitle = {Proceedings of the IEEE/CVF International Conference on Computer Vision},
  pages     = {17443--17453},
  year      = {2025},
  url          ={https://doi.org/10.48550/arXiv.2412.02592}
}

@article{song2026transforming,
  title     = {{Transforming Historical Newspaper Research and Preservation Through AI: A Global Perspective}},
  author    = {Song, Zhao Xun and Cheung, Kwok Wai and Jia, Zi Yun},
  journal   = {Journalism and Media},
  pages     = {10},
  year      = {2026},
  publisher = {MDPI},
  doi = {10.3390/journalmedia7010010}
}

@inproceedings{levchenko2025evaluating,
  title     = {{Evaluating LLMs for historical document OCR: A methodological framework for digital humanities}},
  author    = {Levchenko, Maria A},
  booktitle = {Proceedings of the First on Natural Language Processing and Language Models for Digital Humanities},
  pages     = {75--85},
  year      = {2025},
  doi       = {10.26615/978-954-452-106-6-007}
}

@misc{vesalainen_error_2026,
  title      = {Error {Patterns} in {Historical} {OCR}: {A} {Comparative} {Analysis} of {TrOCR} and a {Vision}-{Language} {Model}},
  url        = {http://arxiv.org/abs/2602.14524},
  doi        = {10.48550/arXiv.2602.14524},
  urldate    = {2026-02-27},
  publisher  = {arXiv},
  author     = {Vesalainen, Ari and Mäkelä, Eetu and Ruotsalainen, Laura and Tolonen, Mikko},
  year       = {2026}
}

@inproceedings{10.1117/12.3104502,
author = {Ying Cao},
title = {{Low-resource adaptive methods for English optical character recognition and quality enhancement using large language models}},
booktitle = {Third International Conference on Big Data, Computational Intelligence, and Applications},
publisher = {SPIE},
pages = {141283U},
year = {2026},
doi = {10.1117/12.3104502},
URL = {https://doi.org/10.1117/12.3104502}
}

@inproceedings{10.1117/12.3068186,
author = {Zhouzhen Shi and Yi Chen},
title = {{Correction of OCR results using large language models}},
booktitle = {International Conference on Advances in Computer Vision Research and Applications (ACVRA)},
year = {2025},
doi = {10.1117/12.3068186},
URL = {https://doi.org/10.1117/12.3068186},
}

@inproceedings{10.1145/3539618.3592057,
author = {Ward, Austin and Avula, Sandeep and Cheng, Hao-Fei and Sarwar, Sheikh Muhammad and Murdock, Vanessa and Agichtein, Eugene},
title = {Searching for Products in Virtual Reality: Understanding the Impact of Context and Result Presentation on User Experience},
year = {2023},
isbn = {9781450394086},
url = {https://doi.org/10.1145/3539618.3592057},
doi = {10.1145/3539618.3592057},
booktitle = {Proceedings of the 46th International ACM SIGIR Conference on Research and Development in Information Retrieval},
pages = {2359–2363},
}

@inproceedings{10.1145/3706599.3720092,
author = {Martins, Tiago and Rom\~{a}o, Teresa},
pages={1–7},
title = {Insights Into the Usability of a Digital Art Archive in Mixed Reality},
year = {2025},
isbn = {9798400713958},
url = {https://doi.org/10.1145/3706599.3720092},
doi = {10.1145/3706599.3720092},
booktitle = {Proceedings of the Extended Abstracts of the CHI Conference on Human Factors in Computing Systems},
articleno = {343},
numpages = {7},
location = {
},
series = {CHI EA}
}

@article{ELMI2026104861,
title = {{H-ProtoRAG: A hierarchical prototype-based retrieval-augmented framework for multilingual CPC patent classification and prior-art retrieval}},
journal = {Information Processing \& Management},
volume = {63},
number = {7, Part B},
pages = {104861},
year = {2026},
issn = {0306-4573},
doi = {10.1016/j.ipm.2026.104861},
url = {https://www.sciencedirect.com/science/article/pii/S0306457326002529},
author = {Zahra Elmi}
}

@article{RUAN2026104784,
title = {A hierarchical RL model for Chinese medical NER with radical semantics and dynamic decision},
journal = {Information Processing \& Management},
year = {2026},
issn = {0306-4573},
doi = {10.1016/j.ipm.2026.104784},
url = {https://www.sciencedirect.com/science/article/pii/S0306457326001755},
author = {Guangce Ruan and Lei Xia}
}

@article{kumpulainen2022struggling,
   title = {Struggling with digitized historical newspapers: Contextual barriers to information interaction in history research activities},
   author = {Sanna Kumpulainen and Elina Late},
   doi = {10.1002/ASI.24608},
   issn = {2330-1643},
   issue = {7},
   journal = {Journal of the Association for Information Science and Technology},
   month = {7},
   pages = {1012-1024},
   publisher = {John Wiley \& Sons, Ltd},
   volume = {73},
   url = {https://onlinelibrary.wiley.com/doi/full/10.1002/asi.24608 https://onlinelibrary.wiley.com/doi/abs/10.1002/asi.24608 https://asistdl.onlinelibrary.wiley.com/doi/10.1002/asi.24608},
   year = {2022}
}

@article{oberbichler2022integrated,
  title   = {Integrated interdisciplinary workflows for research on historical newspapers: Perspectives from humanities scholars, computer scientists, and librarians},
  author  = {Oberbichler, Sarah and Boro{\c{s}}, Emanuela and Doucet, Antoine and Marjanen, Jani and Pfanzelter, Eva and Rautiainen, Juha and Toivonen, Hannu and Tolonen, Mikko},
  journal = {Journal of the Association for Information Science and Technology},
  volume  = {73},
  number  = {2},
  pages   = {225--239},
  year    = {2022},
  doi     = {10.1002/asi.24565}
}

@article{TANG2026104830,
title = {Veritas: Structuring and verifying LLM knowledge for logically consistent behavior tree generation in LLM-based agents},
journal = {Information Processing \& Management},
pages = {104830},
year = {2026},
issn = {0306-4573},
doi = {10.1016/j.ipm.2026.104830},
url = {https://www.sciencedirect.com/science/article/pii/S0306457326002219},
author = {Hao Tang and Feng Zhang and Kejia Wan and Songyi Lu and Xi Zhang and Yiping Yao and Xinhai Xu}
}

@ARTICLE{Zhu_et_al,
  author={Zhu, Xinhua and Wu, Han and Zhang, Lanfang},
  journal={IEEE Transactions on Learning Technologies}, 
  title={Automatic Short-Answer Grading via BERT-Based Deep Neural Networks}, 
  year={2022},
  pages={364-375},
  doi={10.1109/TLT.2022.3175537}}
\end{document}